\documentclass[onecolumn, draftclsnofoot,11pt]{IEEEtran}
\IEEEoverridecommandlockouts
\usepackage{floatflt}
\usepackage{wrapfig}
\usepackage{bbm}
\usepackage{graphics}
\usepackage{amsthm}
\usepackage{amssymb}
\usepackage{amsmath}
\usepackage{graphicx}
\usepackage[colorlinks=true, allcolors=red]{hyperref}
\usepackage{cite}

\usepackage{hyperref}
\hypersetup{
    colorlinks=true,
    linkcolor=black,
    urlcolor=black,
    linktoc=all,
    citecolor=black
}

\usepackage{texPreamble}
\usepackage{texJamesPreamble}

\usepackage{geometry}
\IEEEoverridecommandlockouts

\usepackage{dsfont}

\newtheorem{prop}{Proposition}

\newcommand{\DHy}{D_{\textnormal{H}}}

\newcommand*{\trace}{\mathrm{tr}}
\newcommand*{\set}[1]{\mathcal{#1}}
\newcommand*{\ket}{\rangle}
\newcommand*{\bra}{\langle}

\newcommand{\mcs}{\mathcal{S}}
\newcommand{\mcg}{\mathcal{G}}

\begin{document}

\title{\huge Abelian Group Codes for Classical and Classical-Quantum Channels: One-shot and Asymptotic Rate Bounds}
\author{
% %%% Several authors with up to three affiliations:
\IEEEauthorblockN{James (Chin-Jen) Pang\IEEEauthorrefmark{1},  S. Sandeep Pradhan\IEEEauthorrefmark{2} and Hessam Mahdavifar\IEEEauthorrefmark{3}}\\
       \IEEEauthorrefmark{1}  Marvell Technology, Inc., Santa Clara, CA, USA,       
       \\
      \IEEEauthorrefmark{2}   Department of EECS, University of Michigan, Ann Arbor\\
 \IEEEauthorrefmark{3}   Department of ECE,
                    Northeastern University, Boston, USA, \\
 \IEEEauthorrefmark{1}cjpang@umich.edu 
 \IEEEauthorrefmark{2}pradhanv@umich.edu
 \IEEEauthorrefmark{3}h.mahdavifar@northeastern.edu
\thanks{This paper is presented in part at IEEE International Symposium on Information Theory (ISIT) 2024.}}

\maketitle
\vspace{-0.7in}
\begin{abstract} 
We study the problem of  transmission of information  over classical and classical-quantum channels in the one-shot regime where the underlying codes are constrained to be \emph{group codes}. 
In the achievability part, we introduce a new input probability distribution that incorporates the encoding homomorphism and the underlying channel law. Using a random coding argument, we characterize the performance of group codes in terms of hypothesis testing relative-entropic quantities.
In the converse part, we establish bounds by leveraging a hypothesis testing-based approach.
Furthermore,  we apply the one-shot result to the asymptotic  stationary memoryless setting, and establish a single-letter lower bound  on  \textit{group capacities} for both classes of channels.
Moreover, we derive a  matching  upper bound on the asymptotic group capacity. 
\end{abstract}

\section{Introduction}\label{sec:intro_group}

We study in this paper the problem of channel coding both in the classical  and \emph{classical-quantum} settings. In both problems, the data to be transmitted reliably are classical, but the channel output of the former must have a classical alphabet.
The classical channel coding has been studied extensively in the literature \cite{csiszar2011information}. 
Classical-quantum channel coding has also been studied extensively in a scenario where the channel can be used arbitrarily many times.  
The {channel coding theorem for stationary memoryless classical-quantum channels}, established by Holevo~\cite{holevo1998capacity} and Schumacher and Westmoreland~\cite{schumacher1997sending}, provides an explicit formula for the maximum rate at which data can be transmitted reliably.
% under the assumption that each use of the channel is independent of the previous uses. 
More general channel coding theorems that do not rely on this independence assumption have been developed in later works by Hayashi and Nagaoka~\cite{hayashi2003general} and by Kretschmann and Werner~\cite{kretschmann2005quantum}.
These results are asymptotic, i.e., they refer to a limit where the number of channel uses tends to infinity while the probability of error is required to approach zero. Moreover, the asymptotic performance limits are evaluated for ensembles of  random codes with no apparent structure. We refer to them as random unstructured codes.

Due to its algebraic properties, the finite field structure has been adopted in the coding schemes, over the past several decades, to approach information-theoretic performance limits of point-to-point communication \cite{ahlswede1971bounds, ahlswede1971bounds2, goblick1963coding,dobrusin1962asymptotic, dobrushin1963asymptotic}. These are structured codes where the structure is exploited to yield computationally efficient encoding and decoding algorithms.
Later these coding approaches were extended to weaker algebraic structures such as rings and groups to yield group codes
\cite{forney1993dynamics,como2009capacity,loeliger1991signal,loeliger1996convolutional,garello1995multilevel,sahebi2015abelian, slepian1968group}.
There are several reasons for studying group codes. 
We mention the following two reasons: a) finite fields exist only for alphabets with a prime power size, and b) for communication under certain constraints, codes with weaker algebraic structures have better properties.
For example, when communicating over an additive white Gaussian noise channel with $8$-PSK constellation, codes over $\mathds{Z}_8$, the cyclic group of size $8$, are more desirable over binary linear codes because the structure of the code is matched to the structure of the signal set \cite{loeliger1991signal}. 
As another example, construction of polar codes over alphabets of size $p^r$,  for $r>1$ and $p$ prime,  is simpler with a module structure rather than a vector space structure \cite{sahebi2011multilevel,csacsouglu2009polarization,park2012polar} leading to multilevel polarization. 
In fact, polarization does not take place in finite fields in the polar codes with additive kernel when the field is of size $p^r$ with $r>1$, since components of the finite field remain disconnected in the encoding process. 
Group codes were first studied by Slepian \cite{slepian1968group} for the Gaussian channel. In \cite{ahlswede1971group}, the capacity of group codes for certain classes of channels has been computed. Further results on the capacity of group codes were established in \cite{ahlswede1971bounds, ahlswede1971bounds2}.
The capacity of group codes over a class of channels exhibiting symmetries with respect to the action of a finite Abelian group has been investigated in \cite{como2009capacity}. The capacity of these codes for the discrete memoryless channels was characterized in \cite{sahebi2015abelian}.

Even when computational complexity is a non-issue, structured codes provide performance gains over unstructured codes. 
For example, the typical performance of the ensemble of random structured (linear) codes is better than that of the ensemble of random unstructured codes for point-to-point channels in terms of error exponents and minimum distance \cite{barg2002random,como2009average}. 
These gains get accentuated in network communication setting where structured codes provide gains in the first order 
performance limits such as capacity or rate-distortion functions. 
For example, in problems 
such as distributed source coding, multiple-access channels, interference channels and broadcast channels 
\cite{philosof2009loss, korner1979encode, nazer2007computation, cadambe2008interference, 198101TIT_HanKob, 201408TIT_HonCai, 201407TIT_KriJaf,200809Allerton_SriJafVisJafSha, krithivasan2011distributed,padakandla,hong2014interference}, the use of structured codes have been studied extensively to yield the state-of-the-art bounds on the performance limits. 
Similar first order gains have also been reported in the classical-quantum settings \cite{hayashi2021computation,sohail2022computing,padakandla2022communicating,yao2023capacity} recently.
In these settings, the algebraic structure of the codes are exploited to affect distributed 
coordination among the terminals of the network to yield maximum throughput.

All the aforementioned performance limits are evaluated under the asymptotic assumption, and many assume additionally that the channels are memoryless and stationary. 
However, in many real-world scenarios, we encounter channels which are neither stationary nor memoryless. 
Therefore, it is of fundamental importance to think of coding schemes for the channels which fail to satisfy these assumptions. 
The independent channel uses are relaxed in \cite{han1993approximation,hayashi2003general} and general channels with memory are studied in \cite{datta2007coding, dorlas2011invalidity}, albeit these results are derived in the form of a limit such that the error probability vanishes as the number of channel uses goes to infinity. Hence, the study of performance of codes in the non-asymptotic regimes is of great interest.

In this setting, the problems with  single-serving scenarios, called the one-shot approach,  where a given channel is used only once has been studied extensively in the recent past.  This approach gives rise to a high level of generality that no assumptions are made on the structure of the channel and the associated capacity is usually referred to as \emph{one-shot} capacity.
The one-shot capacity of a classical channel was characterized in terms of min- and max-entropies in \cite{renner2006single}. 
The one-shot classical capacity of a quantum channel is addressed by a hypothesis testing approach in \cite{mosonyi2009generalized} and \cite{wang2012one}, yielding expressions in terms of the generalized (R\'enyi) relative entropies and a smooth relative entropy quantity, respectively. See \cite{toma} for a comprehensive treatment of this approach.
These works considered random unstructured codes--those which do not have any group structure--in their achievability approaches \cite{salek2019one, berta2011quantum, li2021unified, elkayam2020one}. 
In the one-shot approach, the objective is to characterize the performance of a random unstructured code of length unity by using a given channel once. The characterization is given in terms of $\epsilon$ one-shot capacity. The performance of the code on $n$ independent uses of the channel can be obtained by 
evaluating the $\epsilon$-capacity of the $n$-product channel normalized by $n$. Moreover, the characterization should be fine enough that normalized $\epsilon$ capacity for asymptotically large $n$ should give the Shannon capacity of the channel. This approach is also closely related to the second order analysis of capacity of channels as studied in \cite{strassen1962asymptotische,hayashi2009information}.

In this work, we consider the performance of structured codes  for transmission of classical information over point-to-point classical and classical-quantum channels in the one-shot regime. The motivation for this problem is that most codes of practical importance have some form of  algebraic structure, hence it is of significance to obtain a characterization their one-shot coding performance. 
This problem has not been studied before. The first question that arises here is how to formulate the problem. Suppose we want to restrict our attention to just finite fields and formulate the one-shot problem. A code can be defined as  a sub-field of a finite field.  Such a formulation may be overly restrictive. For example, let us take the case of linear codes defined over the binary field $\mathbb{F}_2$. For block length $1$, the alphabet is a finite field, and there is no non-trivial code. 
For any larger block-length $n$, the alphabet is $\mathbb{F}_2^n$. Although one can construct a multiplication operation leading to a Galois field $GF(2^n)$, the standard $(n,k)$ linear code  works only with the algebraic structure of $\mathbb{F}_2$. In fact, it is a subgroup of $\mathbb{F}_2^n$ by considering the latter as just a group, i.e., closed under the addition operation of the base finite field $\mathbb{F}_2$ and, in general, is not a sub-field of $GF(2^n)$. 
This means that we need to go beyond finite fields in our formulation. A similar argument can be given to a formulation 
just based on rings, where a code is a sub-ring of a given ring.
Based on these considerations, we formulate a  one-shot channel coding problem that can encompass a broad class of well-known structured codes.

We assume in this work that the channel input alphabet $G$ is endowed with an Abelian group structure, and we consider a $\mathds{C}$ code characterized via a homomorphism from a group $J$ to $G$, where $J$ is the input message group.
Our derivation is based on the idea of relating the problem of channel coding to hypothesis testing. 
Here, we use a relative-entropy-type quantity defined in \cite{wang2012one} known as hypothesis testing relative entropy, denoted by $\DHy^\epsilon(\cdot\|\cdot)$. 
We introduce a new hypothesis testing group-based  relative entropy that incorporates the underlying subgroup structure of the channel input group alphabet, and  derive a tight characterization of the performance of group codes. 
We use the framework of one-shot quantum typicality developed in  \cite{sen2021unions} for the achievability of multi-terminal CQ channels. The reason for using this multiterminal technique for 
characterizing the group code performance for point-to-point channel is the following. For any non-field group $J$, the lattice \cite{suzuki2012structure} of subgroups induces certain dependence structure on a given  group code, $\mathds{C}$,  bringing with this problem certain issues commonly seen in multi-terminal channel coding problems. Each subgroup of $J$ yields a subcode and one can associate a fictitious terminal with such a subcode. These terminals,  with access to certain parts of the original message,
 form a network induced by the lattice of subgroups.
The overall communication can be interpreted as each such terminal in the network wishing to send a part of the given message with some rate to the receiver.  

The contributions of this paper are as follows. 
We first formulate the one-shot transmission problem with a group structure on the code.  We then provide characterizations of the  one-shot capacities 
of Abelian group codes for classical and classical-quantum channels.
For the former, we assume that the channels have finite alphabets, and for the latter we assume that the channel output Hilbert space is of finite dimension. The achievability is studied using an ensemble of group codes, and we provide two lower bounds on the one-shot capacity, where the latter is tighter while the former is easier to interpret. 
These are stated as Theorem \ref{thm:CC1_achieve}, \ref{thm:CC1_achieve_2nd}, \ref{thm:CQ1_achieve_1}, and \ref{thm:CQ1_achie_2}, respectively. We also provide upper bounds on the one-shot capacities (see Theorem \ref{thm:CC1_conv}, \ref{thm:CQ1_conv}).  These bounds are established by exploiting the hypothesis testing-type argument  on subcodes of $\mathds{C}$ associated with subgroups of $J$.
We then extend these results to the asymptotic regime. 
Using the asymptotic equipartition property (AEP) of the hypothesis testing relative entropy, we rederive a single-letter characterization of the group capacity of discrete stationary and memoryless classical channels which was derived earlier in \cite{sahebi2015abelian} (see Theorem \ref{thm:CC_nshot_cap}).
For the classical-quantum channels, we derive a single-letter characterization of the group capacity in the asymptotic regime using the AEP of hypothesis testing quantum relative entropy under the stationary and memoryless assumption. Furthermore, we provide  a converse coding theorem 
for this capacity (see Theorem \ref{thm:CQ_nshot_cap}).
We note that group codes have been studied extensively in the literature \cite{conway2013sphere}. The asymptotic performance limits of Abelian group codes have been characterized for classical discrete memoryless channels in \cite{como2009capacity,sahebi2015abelian} using varaints of mutual information under maximum likelihood or joint typical decoding procedures.

The new technical elements introduced in this work are as follows. It should be noted that in random ensemble of group codes, introduced in the subsequent sections, the codewords are statistically related, and are not even pairwise independent. 
This requires a new analysis that takes into account the group structure under consideration. One of the key difficulties of working with groups is the presence of zero divisors. 
The second issue that comes up is the decision regions. Because of the lattice of subgroups of a given group, we need to consider intersections of several decision regions. This becomes quite involved in the classical-quantum settings, where the decision regions are the projection operators. 
The third conceptual innovation is the performance characterization. In this regard, we introduce a novel one-shot group hypothesis testing relative entropy toward this.
The paper is organized as follows. In Section \ref{sec:Prelim_group} we provide the basic definitions and the problem formulation. A description of the ensemble of group codes is provided in Section \ref{section:Abelian}.
We characterize the performance of one-shot group channel codes for classical and classical-quantum channels in Section \ref{sec:CC1shot} and \ref{sec:CQ_1shot}, respectively.
Then we derive a single-letter characterization of the asymptotic group capacity of classical and classical-quantum channels in \ref{sec:CCnshot}. We compute the performance limits for some examples in Section \ref{sec:examples}.

\section{Preliminaries}\label{sec:Prelim_group}
\subsection{Classical and CQ Channel Model}

We consider discrete memoryless classical channels used without feedback  specified by a tuple $(\mathcal{X},\mathcal{Y},W_{Y|X})$, where $\mathcal{X}$ and $\mathcal{Y}$ are  the channel input and output alphabets, respectively, and $W_{Y|X}$ is a conditional distribution.
We also study the \emph{classical-quantum channels}, where the data to be transmitted reliably are \emph{classical}. 
Let a finite set $\set{X}$ denote the input alphabet. For any input $x \in \set{X}$, the channel produces an output, specified by a density operator $\rho_x$ on a Hilbert space $\mathcal{B}$. We denote a CQ channel by a mapping $\cqN: x \mapsto \rho_x$ from $\set{X}$ to the set of all density operators defined on $\mathcal{B}$. 

\subsection{Groups and Group Codes}
All groups referred to in this paper are \emph{Abelian groups}. 
Given a group $(G,+)$ and a subset $H$ of $G$, we use $H\le G$ to denote that $H$ is subgroup of $G$.
A \emph{coset} $H'$ of a subgroup $H$ is a shift of $H$ by an arbitrary element $a\in G$, i.e. $H'=a+H$ for some $a\in G$. 
For a prime $p$ dividing the cardinality of $G$, the \emph{Sylow}-$p$ subgroup of $G$ is the largest subgroup of $G$ whose cardinality is a power of $p$.
Group isomorphism is denoted by $\cong$. 
For summations over groups, we also 
use $\sum$.
Direct sum of groups is denoted by
$\bigoplus$ and direct product of sets is denoted by $\bigotimes$.
Given a group $G$, a group code $\mathds{C}$ over $G$ with block length $n$ is any subgroup of $G^n$. A shifted group code over $G$, given by $\mathds{C}+b$ is a translation of a group code $\mathds{C}$ by a fixed vector $b\in G^n$. Group codes generalize the notion of linear codes over primary finite fields to  channels with input alphabets having composite sizes.

\subsection{Definition of Achievability for Classical Channel Coding}\label{subsec:prelim_CCdef}
For a finite Abelian group $G$, a group transmission system with parameters $(n,\Omega,\tau)$ for reliable communication over a given channel $(\mathcal{X}=G,\mathcal{Y},W_{Y|X})$ consists of a codebook, an encoding mapping and a decoding mapping. The codebook $\mathds{C}$ is a shifted subgroup of $G^n$ whose size is equal to $\Omega$ and the mappings are defined as
\begin{align*}
\mbox{\small Enc}:\{1,2,\cdots,\Omega\}\rightarrow \mathds{C}\,, \quad
\mbox{\small Dec}:\mathcal{Y}^n\rightarrow\{1,2,\cdots,\Omega\}\,,
\end{align*}
such that
\begin{align*}
\sum_{m=1}^\Omega \frac{1}{\Omega} \sum_{x\in\mathcal{X}^n} \mathds{1}_{\{x=\mbox{\scriptsize Enc}(m)\}} \sum_{y\in\mathcal{Y}^n} \mathds{1}_{\{m\ne \mbox{\scriptsize Dec}(y)\}} W^n(y|x)\le \tau\, .
\end{align*}
A rate $R$ is said to be achievable using group codes if for all $\epsilon>0$ and for all sufficiently large $n$, there exists a group transmission system with parameters $(n,\Omega,\tau)$ such that
\begin{align*}
\frac{1}{n}\log_2 \Omega \ge R-\epsilon,\qquad\tau\le \epsilon\,.
\end{align*}
The group capacity  $C$ of the channel is defined as the supremum of the set of all achievable rates using group codes.

\subsection{Definition of Achievability for CQ Channel }\label{subsec:prelim_CQdef}

Given a classical-quantum channel $\cqN = \mathset{\rho_x}_{{x\in \cX}} $ from the classical alphabet $\mathcal{X}$ to the quantum system $\mathcal{B}$, where $\mathcal{X}=G$ is an Abelian group,  a group transmission system with parameters $(n,\Omega,\tau)$ over $\cqN$ consists of a codebook, an encoding mapping and a decoding positive operator-valued measure (POVM). The codebook $\mathds{C}$ is a shifted subgroup of $G^n$ whose size is $\Omega$. The encoding mapping is defined as
% \begin{align*}
$\mbox{\small Enc}:\{1,2,\cdots,\Omega\}\rightarrow \mathds{C}.
$
% \end{align*}
The decoding POVM is a set $\mathset{\Lambda_m}_{m=1}^{\Omega}$ of operators such that $\Lambda_m \geq 0, \forall m$ and $\sum_m \Lambda_m = I$. 
The probability of obtaining outcome $j$ is $\trace(\Lambda_j \rho)$ if the state is described by some density operator $\rho$, where $\trace(\cdot)$ denotes the trace operation. 
The probability of decoding error of the group transmission system satisfies
\begin{align*}
\sum_{m=1}^\Omega \frac{1}{\Omega} \sum_{x\in\mathcal{X}^n} \mathds{1}_{\{x=\mbox{\scriptsize Enc}(m)\}} [1-\trace(\Lambda_m \rho_x) ]\le \tau\, .
\end{align*}
Given a channel $\cqN$, a rate $R$ is said to be achievable using group codes if for all $\epsilon>0$ and for all sufficiently large $n$, there exists a group transmission system for reliable communication with parameters $(n,\Omega,\tau)$ such that
\begin{align*}
\frac{1}{n}\log_2 \Omega \ge R-\epsilon,\qquad\tau\le \epsilon\,.
\end{align*}
The group capacity of the channel $C = C(\cqN)$ is defined as the supremum of  all achievable rates using group codes.    

\subsection{Hypothesis testing relative entropy}
We recall the definition and a few properties of the \emph{hypothesis testing relative entropy} $\DHy^\epsilon(\rho\|\sigma)$ from \cite{wang2012one}.
Let $\rho$ and $\sigma$ be two possible states of a system.
 A strategy for the task of discrimination between the two states is specified by a Positive Operator Valued Measure (POVM) with two elements, $\Lambda$ and $I- \Lambda$, corresponding to the two possible values for the guess, respectively, 
 where $\Lambda$ a positive operator with  $0\leq \Lambda \leq I$.
 The probability that the strategy produces a \emph{correct} guess on input $\rho$ is given by $\trace[ \Lambda \rho ]$, and the probability that it produces a \emph{wrong} guess on input $\sigma$ is $\trace[ \Lambda \sigma$]. 
For any $0< \epsilon <1$, the {hypothesis testing relative entropy} $\DHy^\epsilon(\rho\|\sigma)$ is defined by
\begin{align} \label{eq:DHdef}
  \DHy^\epsilon(\rho\|\sigma) \triangleq - \log_2 \inf_{\substack{\Lambda: 0\le
      \Lambda\le I,\\ \trace[\Lambda\rho]\ge 1-\epsilon}} \trace[\Lambda\sigma].
\end{align}
The the following asymptotic equipartition property (AEP) was derived in \cite{tomamichel2013hierarchy,li2014second}:
\[
 \DHy^\epsilon(\rho^{\otimes n} \|\sigma^{\otimes n})=n D(\rho\|\sigma)+\sqrt{n V(\rho\|\sigma)} \Phi^{-1}(\epsilon)+O(\log n),
\]
where $D(\rho\|\sigma)$ is the quantum relative entropy, $V(\rho \|\sigma)$ is the quantum information variance and $\Phi$ is the CDF of the normal distribution.
For two distributions $P,Q$ on a classical alphabet $\cX$, the hypothesis testing relative entropy $D_{\textnormal{H}}^{\epsilon}(P \| Q  )$  can be similarly defined by 
\begin{align} \label{eq:DHdef_classical}
D_{\textnormal{H}}^{\epsilon}(P \| Q ) \triangleq -\log_2 \inf_{A: P(A) \geq 1-\epsilon} Q(A),
\end{align}
where $A\subset \cX$ is sometimes called the \emph{decision region}. Of course,  a similar second order AEP holds here as well.

\section{Abelian Group Code Ensemble}\label{section:Abelian}
In this section, we use the standard characterization of Abelian groups and introduce the ensemble of Abelian group codes used in \cite{como2009capacity,sahebi2015abelian}. The subsequent analysis will be based on this ensemble.
% and this paper.  In other words, we use the standard code ensembles of group codes for our analysis.

\subsection{Abelian Groups}\label{subsec:AbelianGroup}
For an Abelian group $G$, let $\mathcal{P}(G)$ denote the set of all distinct primes which divide $|G|$ and for a prime $p\in\mathcal{P}(G)$ let $S_p(G)$ be the corresponding Sylow subgroup of $G$. It is known that any Abelian group $G$ can be decomposed as a direct sum of its Sylow subgroups \cite{suzuki2012structure} in the following manner
 \begin{align}\label{eqn:G_decomposition}
G=\bigoplus_{p\in \mathcal{P}(G)} S_p(G).
 \end{align}
Furthermore, each Sylow subgroup $S_p(G)$ can be decomposed into $\mathds{Z}_{p^r}$ groups as follows:
$S_p(G)\cong \bigoplus_{r\in\mathcal{R}_p(G)} \mathds{Z}_{p^r}^{M_{p,r}}, $
\begin{align}\label{eqn:Sp_decomposition}
S_p(G)\cong \bigoplus_{r\in\mathcal{R}_p(G)} \mathds{Z}_{p^r}^{M_{p,r}},
\end{align}
where $\mathcal{R}_p(G)\subseteq \mathds{Z}^+$ and for $r\in\mathcal{R}_p(G)$, $M_{p,r}$ is a positive integer. 
Note that $\mathds{Z}_{p^r}^{M_{p,r}}$ is defined as the direct sum of the ring $\mathds{Z}_{p^r}$ with itself for $M_{p,r}$ times. 
Rewriting \eqref{eqn:G_decomposition}, we have:
\begin{align}\label{eqn:G}
G\cong \bigoplus_{p\in\mathcal{P}(G)} \bigoplus_{r\in\mathcal{R}_p(G)} \mathds{Z}_{p^r}^{M_{p,r}}= \bigoplus_{p\in\mathcal{P}(G)} \bigoplus_{r\in\mathcal{R}_p(G)} \bigoplus_{m=1}^{M_{p,r}} \mathds{Z}_{p^r}^{(m)},
\end{align}
where $\mathds{Z}_{p^r}^{(m)}$ is called the $m\textsuperscript{th}$ $\mathds{Z}_{p^r}$ ring of $G$ or the $(p,r,m)$-th ring of $G$. 
We also define two sets, $\mathcal{Q}(G) \subseteq \Bbb{P}\times \Z^+$ by
\[
    \mathcal{Q}(G)  \triangleq \mathset{(p,r)\mid p\in\mathcal{P}(G),  r\in\mathcal{R}_p(G)},
\]
and $\mathcal{G}(G)\subseteq \Bbb{P}\times \Z^+ \times \Z^+$ by
\[
\mathcal{G}(G) \triangleq \mathset{(p,r,m)\mid (p,r)\in \mathcal{Q}(G), m\in \mathset{1, 2, \ldots, M_{p,r}} }.
\] 
Then, equivalently, $G$ can be written as follows
\begin{align}\label{eqn:G_equiv}
     G\cong  \bigoplus_{(p,r)\in\mathcal{Q}(G)} \mathds{Z}_{p^r}^{M_{p,r}} 
     = \bigoplus_{(p,r,m)\in\mathcal{G}(G)} \mathds{Z}_{p^r}^{(m)}.
\end{align}
Define $\zeta(G):= \sum_{(p,r,m) \in \mathcal{G}(G)} r$, 
the sum of prime powers in the prime factorization of $|G|$. 
 Also define 
 the set $\mathcal{G}^*(G) \triangleq \{(p,r,m,k): (p,r,m) \in \mathcal{G}(G), 1 \leq k \leq r\}$.
Hence any element $a$ of the Abelian group $G$ can be regarded as a vector whose components are indexed by $(p,r,m)\in\mathcal{G}(G)$ and whose $(p,r,m)$-th component $a_{p,r,m}$ takes values from the ring $\mathds{Z}_{p^r}$. 

% \begin{comment}
\begin{example}  \label{ex:groupG}
Let $G=\mathds{Z}_4 \oplus
\mathds{Z}_3\oplus\mathds{Z}_9^2$. Then we have
$\mathcal{P}(G)=\{2,3\}$, $S_2(G)=\mathds{Z}_4$ and
$S_3(G)=\mathds{Z}_3\oplus\mathds{Z}_9^2$, $\mathcal{R}_2(G)=\{2\}$,
$\mathcal{R}_3(G)=\{1,2\}$, $M_{2,2}=1$, $M_{3,1}=1$, $M_{3,2}=2$ and
\begin{align*}
\mathcal{G}(G)&=\{(2,2,1),(3,1,1),(3,2,1),(3,2,2)\} \, ,\\
\mathcal{G}^*(G)&=\{(2,2,1,1),(2,2,1,2),(3,1,1,1),(3,2,1,1), 
 (3,2,1,2),(3,2,2,1),(3,2,2,2)\}.
 \end{align*}
 % $\zeta(G)=7$.
Each element $a$ of $G$ can be represented by a quadruple, $(a_{2,2,1},a_{3,1,1},a_{3,2,1},a_{3,2,2})$ where $a_{2,2,1}\in\mathds{Z}_4$, $a_{3,1,1}\in\mathds{Z}_3$ and $a_{3,2,1},a_{3,2,2}\in\mathds{Z}_9$. 
\end{example}
% \end{comment}
In the following section, we introduce the ensemble of Abelian group codes which we use in the paper.

\subsection{The Image Ensemble}\label{subsec:Image_Ensemble}
Recall that for a positive integer $n$, an Abelian group code of length $n$ over the group $G$ is a coset of a subgroup of $G^n$. Our ensemble of codes consists of \textit{all} Abelian group codes over $G$; i.e., we consider all cosets of subgroups of $G^n$. The following lemma (see \cite[Theorem 12-1]{bloch1987abstract}) effectively characterizes all subgroups of $G^n$:
\begin{lemma} \label{lem:homo_JtoG}
For a group $\tilde{G}$, let $\phi:J\rightarrow \tilde{G}$ be a homomorphism from some group $J$ to $\tilde{G}$. Then $\phi(J)\le \tilde{G}$. Moreover, for any subgroup $\tilde{H}$ of $\tilde{G}$ there exists a corresponding group $J$ and a homomorphism $\phi:J\rightarrow \tilde{G}$ such that $\tilde{H}=\phi(J)$.
\end{lemma}

\begin{definition}\label{def:x_decomp}
    For an element $x \in G$ and a subgroup $H \leq G$, there is a one-to-one mapping $x \leftrightarrow ([x], \overline{x})$, where $[x]$ is a representative of the coset of $H$ which $x$ belongs to, and $\overline{x} \in H$, such that $x =[x] + \overline{x}$. 
\end{definition}

\begin{remark}
\label{rem:dependency}
We make a general remark here about an arbitrary subgroup $H$ of $G$.
Let $\phi:J\rightarrow {G}$ be a homomorphism from some group $J$ to ${G}$, and $x = \phi(u)+\beta$ for any fixed $\beta \in G$. Let $H_J=\phi^{-1}(H)$, and 
let $u=[u]+\overline{u}$ be the decomposition as in Definition \ref{def:x_decomp} with respect to $H_J$. 
We note that $[x]= x \mod H$, and 
$\phi(\overline{u})=\overline{x}$. Using the distributive property of $\!\!\!\! \mod \!\!$ operation, 
i.e., $[x+y]=[[x]+y]$, we note that,
\[
[x]=[\phi(u)+\beta] =[[\phi([u])+\phi(\overline{u})] +\beta] = 
[[\phi([u])]+\beta], 
\]
\[
\overline{x}=x-[x]=\phi(\overline{u})+
\overline{\phi([u])+\beta}.
\]
Hence $[x]$ depends only on $[u]$, where as $\overline{x}$ depends on the entirety of $u$ in general.  This is the characteristic feature of the group codes. 
% \blue{ $[u]$ is w.r.t. which the subgroup ? Maybe an example? }
\end{remark}
\begin{definition}\label{def:rp} 
Let $G$ be an Abelian group. For $p\in\mathcal{P}({G})$, define
% \begin{align}\label{eqn:r_p}
$r_{p}=\max \mathcal{R}_p({G})$,
% \end{align} 
% $\mathcal{S}_p(G)=\{1,2,\cdots,r_p\}$, 
and $\mathcal{S}(G)=\{(p,s)\mid p\in\mathcal{P}(G), 1\leq s\leq r_{p}\}.$
\end{definition}
It is shown in \cite{sahebi2015abelian} that we only need to consider homomorphisms from an Abelian group $J$ to $\tilde{G}$ such that 
% \begin{align*}
$\mathcal{P}(J)\subseteq \mathcal{P}(\tilde{G}), \mbox{ and }
 s\le r_q = \max \mathcal{R}_q(\tilde{G})  
  \mbox{ for all } (q,s,l) \in\mathcal{G}(J).
$
% \end{align*}
To construct Abelian group codes of length $n$ over $G$, we start with the following decomposition:
\begin{align}\label{eqn:Gn}
G^n&\cong\bigoplus_{p\in\mathcal{P}(G)} \bigoplus_{r\in\mathcal{R}_p} \mathds{Z}_{p^r}^{n M_{p,r}} =\bigoplus_{p\in\mathcal{P}(G)} \bigoplus_{r\in\mathcal{R}_p} \bigoplus_{m=1}^{n M_{p,r}} \mathds{Z}_{p^r}^{(m)}\, .
% =\bigoplus_{(p,r,m)\in\mathcal{G}(G^n)} \mathds{Z}_{p^r}^{(m)}
\end{align}
Define $J$ as
\begin{align}\label{eqn:J}
J=\bigoplus_{q\in\mathcal{P}(G)} \bigoplus_{s=1}^{r_q} \mathds{Z}_{q^s}^{k_{q,s}} =\bigoplus_{q\in\mathcal{P}(G)} \bigoplus_{s=1}^{r_q} \bigoplus_{l=1}^{k_{q,s}} \mathds{Z}_{q^s}^{(l)},
% =\bigoplus_{(q,s,l)\in\mathcal{G}(J)} \mathds{Z}_{q^s}^{(l)}
\end{align}
for some non-negative integers $k_{q,s}$.
Define
% \begin{align*}
$k=\sum_{q\in \mathcal{P}(G)} \sum_{s=1}^{r_q} k_{q,s}$
% \end{align*}
and $w_{q,s}=\frac{k_{q,s}}{k}$ for $(q, s) \in \mathcal{S}(G)$. 
Any homomorphism $\phi:J\rightarrow G^n$ can be represented by 
\begin{align}\label{eqn:phi}
\phi(a)=\bigoplus_{(p,r,m)\in\mathcal{G}(G^n)} {\sum}_{(q,s,l)\in\mathcal{G}(J)} a_{q,s,l} g_{(q,s,l)\rightarrow (p,r,m)}, 
\end{align}
for $a\in J$, 
where $a_{q,s,l} g_{(q,s,l)\rightarrow (p,r,m)}$ is the short-handed notation for the mod-$p^r$ addition of $g_{(q,s,l)\rightarrow (p,r,m)}$ to itself for $a_{q,s,l}$ times and the sum is over $\mathds{Z}_{p^r}$. Further, it is shown in \cite[Lemma 2]{sahebi2015abelian} that we must have 
\begin{equation}\label{eq:g_constraints}
g_{(q,s,l)\rightarrow (p,r,m)} = 0 \mbox{ if } p\neq q\qquad \mbox{ and } \quad 
g_{(q,s,l)\rightarrow (p,r,m)} \in p^{r-s}\Z_{p^r} \mbox{ if } p =q, r\geq s\, .
\end{equation}

\begin{definition}\label{def:groupCode}
The ensemble of Abelian group encoders consists of all mappings $\phi:J\rightarrow G^n$  in the form of \eqref{eqn:phi} % Let $k=\sum_{q\in\mathcal{P}(G)} \sum_{s=1}^{r_q} k_{q,s}$ and define $w_{q,s}=\frac{k_{q,s}}{k}$.
and 
\[
g_{(q,s,l)\rightarrow (p,r,m)}\begin{cases}
    =0 &\mbox{ if } p\ne q\\
    \sim  \textnormal{Unif}(\mathds{Z}_{p^r})  &\mbox{ if }  p=q, r\le s\\
    \sim \textnormal{Unif}(p^{r-s}\mathds{Z}_{p^r})  &\mbox{ if } p=q, r\ge s
\end{cases}
\]
The corresponding group code is defined by
\begin{align}\label{eqn:Code}
\mathds{C}=\{\phi(a)+V|a\in J\},
\end{align}
where $V$ is a uniform random variable over $G^n$. 
\end{definition}
The rate of this code is given by
\begin{align}\label{eqn:rate}
R&=\frac{1}{n}\log |J|=\frac{k}{n}\sum_{q\in\mathcal{P}(G)} \sum_{s=1}^{r_q} s w_{q,s} \log q.
\end{align}

  We can express the code compactly as follows:
\begin{align*}
\mathds{C} &= \Bigg\{ \bigoplus_{i=1}^n \left[ \bigoplus_{(p,r,m)=\mcg(G)} \sum_{s=1}^{r_p} a_{p,s} g^{(i)}_{(p,s) \rightarrow (r,m)} 
+V^{(i)} \right]  
  :  a_{p,s} \in \mathds{Z}_{p^s}^{k w_{p,s}}, \forall
(p,s) \in \mcs(G) \Bigg\},
\end{align*}
where $g^{(i)}_{(p,s) \rightarrow (r,m)} \in p^{|r-s|^+}\mathds{Z}_{p^r}^{k w_{p,s}}$, and the product is given by the component-wise product followed by the addition operation. 

\subsection{The $H_{\hat{\theta}}$ coset}\label{subsec:H_coset}

For an Abelian group $G$ defined in \eqref{eqn:G}, denote a vector $\hat{\theta}$ whose components are non-negative integer-valued and indexed by $(p,s)\in\mathcal{S}(G)$ by $(\hat{\theta}_{p,s})_{(p,s)\in\mathcal{S}(G)}$, where $0\leq \hat{\theta}_{p,s} \leq s$. 
Let $\bs$ denote the vector whose components satisfy $\bs_{(p,s)} = s$ for all $(p,s)\in \mathcal{S}(G)$.
Let $\Theta = \Theta(G)$ be the set of vectors $\hat{\theta}$ indexed by $(p,s) \in \mathcal{S}(G)$ such that $0\leq \hat{\theta}_{p,s} \leq s$ and $\hat{\theta} \neq \bs$, and denote its size by $M\triangleq \abs{\Theta}.$ 
For $a\in J$ and $\hat{\theta}=(\hat{\theta}_{p,s})_{(p,s)\in\mathcal{S}(G)}$, let $T_{\hat{\theta}}(a)$ denote the set of vectors $\tilde{a}\in J$ such that 
\begin{align*}
\tilde{a}_{p,s} - a_{p,s} \in 
p^{\hat{\theta}_{p,s}}\mathds{Z}_{p^s}^{k_{p,s}}
\backslash
p^{\hat{\theta}_{p,s}+1}\mathds{Z}_{p^s}^{k_{p,s}}, \forall (p,s)\in\mathcal{S}(G)\ ,
\end{align*} where we follow the convention and set 
$p^{s}\mathds{Z}_{p^s}^{k_{p,s}}
\backslash
p^{s+1}\mathds{Z}_{p^s}^{k_{p,s}} = \mathset{0}$.
Then we have
$\abs{T_{\hat{\theta}}(a)}
= \prod_{(p,s)\in\mathcal{S}(G)} p^{(s- \hat{\theta}_{p,s}) k_{p,s}}
$ for all $a\in J$. 
Therefore, we may write $\abs{T_{\hat{\theta}}(a)} = \abs{T_{\hat{\theta}}}$ without any ambiguity. 
Let $\omega_{\hat{\theta}}$ be defined by
 \begin{align}\label{eqn:omega_theta}
\omega_{\hat{\theta}} &\triangleq \frac{\sum_{(p,s)\in\mathcal{S}(G)} \hat{\theta}_{p,s} w_{p,s} \log p } {\sum_{(p,s)\in\mathcal{S}(G)} s w_{p,s} \log p}, 
\end{align} we show in Appendix~\ref{appendix:H_coset} the following result: 
\begin{align}
\log \abs{T_{\hat{\theta}}} = (1- \omega_{\hat{\theta}} ) nR\ . \label{eq:T_theta_size}
\end{align}
For any $a\in J$,  $\mathset{{T_{\hat{\theta}}(a)}}_{\hat{\theta}} $ is a union of disjoint sets whose union is $\cup_{\hat{\theta} } T_{\hat{\theta}}(a) = J $. Hence $\sum_{\hat{\theta}}  \abs{T_{\hat{\theta}}} =  \sum_{\hat{\theta}}  \abs{T_{\hat{\theta}}(a)} = \abs{J}$.
Exploiting \eqref{eq:T_theta_size}, we have that $
\sum_{\hat{\theta}} 2^{ {(1- \omega_{\hat{\theta}} ) nR} } = \abs{J}$, or equivalently, $\sum_{\hat{\theta}} 2^{ {(1- \omega_{\hat{\theta}} ) } } =1$.

For $\hat{\theta}=(\hat{\theta}_{p,s})_{(p,s)\in\mathcal{S}(G)}$, define a vector $\pmb{\theta}(\hat{\theta})$ indexed by $(p,r)\in \mathcal{Q}(G)$ and 
\begin{align*}
\left(\pmb{\theta}(\hat{\theta})\right)_{p,r}
\triangleq \min_{\substack{(p,s)\in\mathcal{S}(G)\\w_{p,s}\ne 0}} |r-s|^+ +\hat{\theta}_{p,s}\ .
\end{align*}
Let $H_{\hat{\theta}}$ be a subgroup of $G$ defined as
\begin{align}\label{eqn:H_theta}
H_{\hat{\theta}} \triangleq 
\bigoplus_{(p,r,m)\in\mathcal{G}(G)} p^{\pmb{\theta}(\hat{\theta})_{p,r}} \mathds{Z}_{p^r}^{(m)}.
\end{align}
As shown in Lemma \ref{lem:Pr_Ttheta} in Appendix~\ref{appendix:H_coset}, if $\tilde{a} \in T_{\hat{\theta}}(a)$ for some $a \in J$, the difference $  \phi(\tilde{a})  - \phi( a) =  \phi(\tilde{a} - a)$ is uniformly distributed over $H_{\hat{\theta}}^n$, for the code ensemble described in Definition~\ref{def:groupCode}.

\begin{example}\label{exam:theta_hat}  
    Let $G=\mathbb{Z}_4  \oplus \mathbb{Z}_8$, and $J=\mathbb{Z}_2 \oplus \mathbb{Z}_4 \oplus \mathbb{Z}_8$ and let  $n=1$. Here $r$ takes $2$ values and $s$ takes $3$ values. The set 
    $\mathcal{P}(G)$ is the singleton set of $2$, $r_2 =  3$, and the set
    $\mathcal{S}(G) = \mathset{(2,1), (2,2), (2,3) }$. 
    Also, we have $k_{2,1} = k_{2,2} =k_{2,2} =1$, $k =3$, and  $w_{2,1} = w_{2,2} =w_{2,2} =1/3$.
    A vector $\hat{\theta}$ has three entries $ 0\leq \hat{\theta}_{2,1} \leq 1, 0\leq \hat{\theta}_{2,2} \leq 2, 0\leq \hat{\theta}_{2,3} \leq 3$.
    The vector $\pmb{\theta}(\hat{\theta})$ is indexed by $(p,r)\in \mathcal{Q}(G) = \mathset{(2,2), (2,3)}$, and 
    \begin{align*}
        (\pmb{\theta}(\hat{\theta}))_{2,2} = 
    \min_{ 1\leq s \leq 3} \left(|2-s|^+ +\hat{\theta}_{2,s} \right)\, , \quad  
        (\pmb{\theta}(\hat{\theta}))_{2,3} = 
    \min_{ 1\leq s \leq 3} \left(|3-s|^+ +\hat{\theta}_{3,s}\right)\, . 
    \end{align*} We consider two cases of $\hat{\theta}$ and show the corresponding $\pmb{\theta}(\hat{\theta})$ and $H_{\hat{\theta}}$. 
    First, consider $\hat{\theta} = (0, 0, 0)$. Then  $\pmb{\theta}(\hat{\theta}) = (0, 0)$, and $H_{\hat{\theta}}= 
    \Z_4 \bigoplus \Z_8$.
    For  $a\in J$,  a vector $\tilde{a} \in T_{\hat{\theta}}(a)$ has 
    \begin{align*}
        \tilde{a}_{2,1} - a_{2,1} \in \Z_2 \backslash 2\Z_2 , \quad
        \tilde{a}_{2,2} - a_{2,2} \in \Z_4 \backslash 2\Z_4 , \quad
        \tilde{a}_{2,3} - a_{2,3} \in \Z_8 \backslash 2\Z_8.
    \end{align*} 
    Considering all homomorphisms $\phi$ as in \eqref{eqn:phi} satisfying \eqref{eq:g_constraints}, one can show that the set $\mathset{ \phi(\tilde{a} - a)}_{\tilde{a} \in T_{\hat{\theta}}(a) }$ is equal to $H_{\hat{\theta}}$. 
    For another case, consider $\hat{\theta} = (1, 1, 2)$. Then  $\pmb{\theta}(\hat{\theta}) = (1, 2)$, and $H_{\hat{\theta}}= 
    2\Z_4 \bigoplus 4\Z_8$.
    For  $a\in J$,  a vector $\tilde{a} \in T_{\hat{\theta}}(a)$ has 
    \begin{align*}
        \tilde{a}_{2,1} - a_{2,1} \in 2\Z_2 \backslash 4\Z_2 , \quad
        \tilde{a}_{2,2} - a_{2,2} \in 2\Z_4 \backslash 4\Z_4 , \quad
        \tilde{a}_{2,3} - a_{2,3} \in 4\Z_8 \backslash 8\Z_8 ,
    \end{align*}
    that is $\tilde{a} = ( \tilde{a}_{2,1}, \tilde{a}_{2,2}, \tilde{a}_{2,3}) = ( a_{2,1},  a_{2,2} + 2,  a_{2,3} + 4)$. Similarly, $\mathset{ \phi(\tilde{a} - a)}_{\tilde{a} \in T_{\hat{\theta}}(a) }$ is equal to $H_{\hat{\theta}}$. 
\end{example}

\subsection{Regular versus irregular codes}
\label{sec:regular}

We have seen that for a given pair $(J,G^n)$, any homomorphism between them must be of the form given in \eqref{eqn:phi}.
Every such homomorphism can be characterized using the collection
$\{g^{(i)}_{(p,s) \rightarrow (r,m)}\}$
and a corresponding code is given by 
$\mathbb{C}=\{\phi(a)+v | a \in J\},$
where $v \in G$ is a shift element. 
It is then possible to construct group codes where the image of the homomorphism belongs to 
$\bigoplus_{i=1}^n H_i$, where $H_i \leq G$, and some or all of $H_i$ may be proper subgroups of $G$. To bookkeep such cases, 
consider an arbitrary coordinate  $i$.
One can associate two vectors $\eta$ and $b$ of $G$ for $i$ as follows. 
The components of the vector $\eta$
 are indexed by $(p,r,m,s)$, for every $(p,r,m) \in \mathcal{G}(G)$ and $0 \leq s \leq r_p$, and satisfies
 $0 \leq \eta_{p,r,m,s} \leq r-|r-s|^+$. 
The components of $b$ are indexed by 
$(p,r,m)\in \mathcal{G}(G)$ and
$b_{p,r,m} \in \mathbb{Z}_{p^r}$. 
The components are characterized as follows
\[
g^{(i)}_{(p,s) \rightarrow (r,m)} \in p^{\eta_{p,r,m,s}+|r-s|^+} \mathbb{Z}_{p^r}^{kw_{p,s}}, \quad 
g^{(i)}_{(p,s) \rightarrow (r,m)} \not \in p^{\eta_{p,r,m,s}+1+|r-s|^+} \mathbb{Z}_{p^r}^{kw_{p,s}}, \quad b_{p,r,m}=v_{p,r,m}.
\]
A code is said to be regular if $\eta$ is the all-zero vector for all $i$. If not, the code is said to be irregular. For irregular codes, the image of the homomorphism may be restricted to a coset of a proper subgroup of $G$.

\begin{example}    
    Let $G=\mathbb{Z}_4  \oplus \mathbb{Z}_8$, and $J=\mathbb{Z}_2 \oplus \mathbb{Z}_4 \oplus \mathbb{Z}_8$. 
Here $r$ takes two values and $s$ takes $3$ values.
    The three input components are mapped to the two output components. Consider the case when $\eta$  is the all-zero vector.  
Then $g_{(p,s) \rightarrow (r,m)} \in 2^{r-s} \mathbb{Z}_{2^r} \backslash 2^{r-s+1} \mathbb{Z}_{2^r}$, for $r \geq s$, and 
$g_{(p,s) \rightarrow (r,m)} \in \mathbb{Z}_{2^r} \backslash 2^{1} \mathbb{Z}_{2^r}$ for $r <s$.
The image of the homomorphism is equal to $G$. 
Next consider the case when $\eta$ is the all-ones vector. Then 
$g_{(p,s) \rightarrow (r,m)} \in 2^{r-s+1} \mathbb{Z}_{2^r} \backslash 2^{r-s+2} \mathbb{Z}_{2^r}$, for $r \geq s$, and 
$g_{(p,s) \rightarrow (r,m)} \in 2^{1}\mathbb{Z}_{2^r} \backslash 2^{2} \mathbb{Z}_{2^r}$ for $r <s$. The image of the homomorphism takes values in $2 \mathbb{Z}_4 \oplus 2\mathbb{Z}_8$. 
\end{example}
For $\hat{\theta}$, and $\eta$, define
\begin{align*}
\pmb{\theta}(\eta) \triangleq 
\big(\min_{\substack{1 \leq s \leq r_p\\w_{p,s}\ne
    0}} |r-s|^+ +\eta_{p,r,m,s}\big)_{(p,r,m)\in \mathcal{P}(G)}.
\end{align*}
Define  $H_{\eta} \leq G$ and $H_{\eta+\hat{\theta}} \leq H_{\eta}$ as 
\begin{align}\label{eqn:H_eta}
H_{\eta} \triangleq \bigoplus_{(p,r,m)\in\mathcal{G}(G)}
p^{\pmb{\theta}(\eta)_{p,r,m}} \mathds{Z}_{p^r}^{(m)}, \quad 
H_{\eta+\hat{\theta}} \triangleq \bigoplus_{(p,r,m)\in\mathcal{G}(G)}
p^{\pmb{\theta}(\eta+\hat{\theta})_{p,r,m}} \mathds{Z}_{p^r}^{(m)},
\end{align}
where $\eta+\hat{\theta}$ denote a vector obtained by adding
the components with index $(p,s)$ in $\mcs(G)$.
The image of the corresponding homomorphism is equal to $H_{\eta}$.

We use the following notations for the conditional distributions of the codeword and channel output given the \textit{coset} information. 
\begin{definition}\label{def:Htheta_cond_CC}
For any pair $(\eta,b)$ and any vector $\hat{\theta}$,
 let $X$ be distributed according to $P_X \equiv  \textnormal{Unif}(H_{\eta}+b)$, the uniform distribution over $H_{\eta}+b$. 
Then, for a representative   $\sparenv{x_r}$ of a coset of $H_{\eta+\hat{\theta}}$ in $H_{\eta}+b$, define 
\begin{align*}
P_{[X]}(\sparenv{x}) &\triangleq \Pr([X] = [x]) = \frac{\abs{ H_{\eta+\hat{\theta}}}}{\abs{H_{\eta}}},  \\
P_{X\mid [X]}(x \,\vert \sparenv{x_r}) &\triangleq \Pr(X =x \mid [X] = [x_r])  
=  \Big\{ \begin{array}{cc} \frac{1}{\abs{{H_{\eta+\hat{\theta}}}}} 
 & \mbox{ if } x \in [x_r],    \\
 0 &  \mbox{ otherwise,}  \end{array} 
\end{align*}
where we use $ x \in [x_r]$ as a shorthand for $  x \in [x_r] + H_{\eta+\hat{\theta}}$, when it is clear from the context.
When a classical channel  $(\cX, \cY, W_{Y\mid X})$ is given, we define
\begin{align*}
P_{Y\mid [X]}(y\,\vert  \sparenv{x_r}) &\triangleq 
\Pr(Y = y \mid [X] = [x_r]) 
=\sum_{x\in [x_r] } P_{X\mid [X]}(x| [x_r]) W_{Y\mid X}(y| x)  
 = \sum_{x\in \sparenv{x_r} } \frac{1}{\abs{H_{\eta+\hat{\theta}}}} W_{Y|X}(y\vert x).
\end{align*}
\end{definition}
An analogous  quantum object will be defined in the case of classical-quantum channel in Section \ref{sec:CQ_1shot}.

\section{One-shot Classical Group Coding}
\label{sec:CC1shot}
Given a channel $(\mathcal{X}=G,\mathcal{Y},W_{Y|X})$, and a pair $(\eta,b)$,  let the joint distribution $P_{X_{\eta,b}Y}$ be $P_{X_{\eta,b}Y} = P_{X_{\eta,b}} \cdot W_{Y|X}$ with $P_{X_{\eta,b}}$ being the uniform distribution over $H_{\eta}+b$, and let $P_Y$ denote the marginal distribution of $P_{X_{\eta,b}Y}$ over $\cY$.
Consider a vector $\hat{\theta} \in \Theta(G)$. Let $H_{\eta+\hat{\theta}}$ be a subgroup of $G$ defined in \eqref{eqn:H_eta} and  let 
$[X_{\eta,b}]_{\hat{\theta}}$ be the  representative 
of the coset of $H_{\eta+\hat{\theta}}$ to which $X_{\eta,b}$ belongs. Moreover\footnote{$\hat{\theta}$ is not explicit in the notation $\overline{X}$ to reduce clutter. It should be clear from the context.}, $\overline{X}_{\eta,b}=X_{\eta,b}-[X_{\eta,b}]_{\hat{\theta}}$.
Define
\begin{align}
% \begin{split}
    &I_{\textnormal{H}}^{\epsilon}(\overline{X}_{\eta,b}; [X_{\eta,b}],Y) \triangleq D_{\textnormal{H}}^{\epsilon}(P_{X_{\eta,b} Y} \| P_{[X_{\eta,b}]} P_{X_{\eta,b}\mid [X_{\eta,b}]} P_{Y\mid [X_{\eta,b}]} ). \label{eq:DH_coset_def}
\end{align}
We can also denote this quantity as $I_{\textnormal{H}}^{\epsilon,\hat{\theta}}(\overline{X}_{\eta,b}; [X]_{\eta,b}Y)$ if we would like to make its dependence on $\hat{\theta}$ explicit.
When $\eta$ and $b$ are clear from the context, this quantity is also denoted as 
$I_{\textnormal{H}}^{\epsilon,\hat{\theta}}(\overline{X}; [X]Y)$ or $I_{\textnormal{H}}^{\epsilon}(\overline{X}; [X]Y)$. 

\subsection{Achievability}\label{subsec:CC1shot_achieve}
We have the first main result of this section. This characterizes an achievable one-shot performance of group codes
on a classical channel. 
\begin{theorem}\label{thm:CC1_achieve}
Let $\epsilon$ and $\mathset{\epsilon_{\hat{\theta}}} $ be given with $ \epsilon_{\hat{\theta}}>0$ for all $\hat{\theta}$, and  $\sum_{\hat{\theta}\neq \bs}  \epsilon_{\hat{\theta}} \leq \epsilon$. 
For a given classical channel $(G,\mathcal{Y},W_{Y|X})$, any $(\eta,b)$, and any $J$ as given in \eqref{eqn:J},
there exists a $(1, \abs{J}, \epsilon')$- group transmission system such that 
\[
\epsilon' \leq  \epsilon + \sum_{{\hat{\theta}} \neq \bs }   \exp_2 \bracenv{ 
(1- \omega_{\hat{\theta}} )R
 -I_{\textnormal{H}}^{\epsilon_{\hat{\theta}}}(\overline{X}_{\eta,b}; [X_{\eta,b}],Y) },\]
where the rate $R$ is given in \eqref{eqn:rate}.
\end{theorem}

\begin{remark}
    This theorem motivates the definition of one-shot $\epsilon$-group capacity as
    \[
    \max_{\eta,b} \min_{\hat{\theta} \neq \mathbf{s}}  \frac{1}{(1-w_{\hat{\theta}})} 
    I_{\textnormal{H}}^{\epsilon_{\hat{\theta}}}(\overline{X}_{\eta,b}; [X_{\eta,b}],Y).
    \]
    We compute this quantity for several examples in Section \ref{sec:examples}.
    First consider the case when $\eta =0$, and $b$ is arbitrary. 
    Note that there is a constraint on the rate for every subgroup of the input group $J$, indexed by $\hat{\theta}$. 
    The constraint can be interpreted as follows. Suppose a message $a \in J$ is sent by the encoder. Then  
    $\exp_2[-I_{\textnormal{H}}^{\epsilon_{\hat{\theta}}}(\overline{X}_{\eta,b}; [X_{\eta,b}],Y)]$ corresponds to the probability of the event that some other message $\tilde{a}$ looks statistically related to the output $Y$, and $\tilde{a} \in  T_{\hat{\theta}}(a)$, i.e., they belong to the same coset of the subgroup. The number of such $\tilde{a}$ is given by 
$\exp_2[(1-w_{\hat{\theta}})R]$. This set of constraints is the price to be paid for endowing the code with the group structure. This is also the penalty for the loss of pairwise independence of the codewords in the group ensemble. This will be evident when we discuss  the proof of the theorem. 
\end{remark}

\begin{remark}
    When $G$ is a simple group, then there is no non-trivial subgroup and the rate is given by  $I_{\textnormal{H}}^{\epsilon}(X;Y)$.   This corresponds to the case when $G$ is a finite field.  This is also the one-shot $\epsilon$-capacity achieved by the random unstructured code ensemble with uniform input distribution. Then the group structure comes for free.  However, for the unstructured code ensemble, one has the freedom to use any non-uniform distribution on the input alphabet, which is forbidden for group codes defined in terms of group homomorphisms.
    The reason for this is that when the distribution on the input group $J$ is uniform, i.e.,  message distribution is uniform, then the single-letter output distribution will be uniform on a coset of a subgroup of $G$ due to the homomorphic nature of the encoding. In a group code ensemble, there is less freedom in choice of a codeword assigned to a particular message due to the global algebraic structure of the code. 
    For any non-trivial $\eta,b$, the image of the code takes values in a coset of a proper subgroup of $G$.
\end{remark}

\noindent \textbf{Proof:}   
We give a detailed proof for the case when $\eta=\zero$, and $b$
 is arbitrary. 
 An extension to any general $\eta$ vector is straightforward by redefining the group $G$ accordingly as $H_{\eta}$, and by redefining the channel as first adding $b$ to the input and then 
 acting with random transformation given by $W_{Y|X}$.

Let the ensemble of homomorphisms $\phi$ from $J$ to $G$  and the group code $\mathds{C}=\{\phi(a)+V|a\in J\}$ be given as in Definition~\ref{def:groupCode} with $n=1$. 
% \end{definition}
% Given a channel $W =(\mathcal{X}=G,\mathcal{Y},W_{Y|X})$, 
Let a set of parameters $\mathset{\epsilon_{\hat{\theta}}}_{{\hat{\theta}} \neq \bs}$ be given such that  ${\epsilon_{\hat{\theta}}} >0$ for each $\hat{\theta}$ and that $\sum_{\hat{\theta}} \epsilon_{\hat{\theta}} = \epsilon$. 
Consider a decision region $A_{\epsilon} \subset \cX\times \cY$, which will be constructed explicitly in the sequel, such that 
\begin{equation}
    P_{XY}( A_{\epsilon}) = \sum_{(x,y)\in A_{\epsilon}} P_X(x) W_{Y|X}(y\vert x) \geq 1-\epsilon,
\end{equation}  where $P_X$ is uniform over $G$.

To find an achievable rate, we use a random coding argument in which the random encoder is characterized by the random homomorphism $\phi$ and a random vector $V$ uniformly distributed over $G$. Given a message $u\in J$, the encoder maps it to $x=\phi(u)+V$ and $x$ is then fed to the channel. At the receiver, after receiving the channel output $y\in\mathcal{Y}$, the decoder looks for a unique $\tilde{u}\in J$ such that $(\phi(\tilde{u})+V,y) \in A_{\epsilon}$. If the decoder does not find such $\tilde{u}$ or if such $\tilde{u}$ is not unique, it declares error.
% Let $u$, $x$ and $y$ be the message, the channel input and the channel output respectively.
Thus, the error event can be characterized by the union of two events: $E(u)=E_1(u)\cup E_2(u)$ where $E_1(u)$ is the event that $(\phi(u)+V, y)\notin A_{\epsilon}$ and $E_2(u)$ is the event that there exists a $\tilde{u}\ne u$ such that $(\phi(\tilde{u})+V,y) \in A_{\epsilon}$. 
We can provide an upper bound on the probability of the error event as 
\[
\Pr(E(u))\le \Pr(E_1(u))+\Pr(E_2(u)\cap (E_1(u))^c).
\]
Now we specify the region $A_{\epsilon}$. Let the set $A^{\ast}_{\epsilon_{\hat{\theta}}}$ be a minimizer in the definition of the right-hand side of \eqref{eq:DHdef_classical} for 
$I_{\textnormal{H}}^{\epsilon_{\hat{\theta}}}(\overline{X}; [X]Y),$
% $D_H^{\epsilon_{\hat{\theta}}, \hat{\theta}} (P_{XY} \| P_{[X]} P_{X\mid [X]} P_{Y\mid [X]} )$
 i.e., $P_{XY}(A^{\ast}_{\epsilon_{\hat{\theta}}}) \geq 1- \epsilon_{\hat{\theta}}$ and
\begin{align*}
    I_{\textnormal{H}}^{\epsilon_{\hat{\theta}}}(\overline{X}; [X]Y)
    = -\log_2  
\Big[   \sum_{\sparenv{x_r}} P(\sparenv{x_r}) 
    \sum_{x\in \sparenv{x_r}} P(x \,\vert \sparenv{x_r})
    \sum_{y: (x,y)\in A^{\ast}_{\epsilon_{\hat{\theta}}}}
    P(y\,\vert  \sparenv{x_r})\Big]\, .
\end{align*}
Define the region $A_{\epsilon}$ as $A_{\epsilon} = \cap_{\hat{\theta} \neq \bs} A^{\ast}_{\epsilon_{\hat{\theta}}} $.  
In Appendix \ref{app:CC_Error}, we show that  
 \begin{align}
 \Pr(E(u)) \leq \epsilon + \sum_{{\hat{\theta}} \neq \bs }  \abs{T_{{\hat{\theta}}}(u)} \exp_2 \bracenv{ -
I_{\textnormal{H}}^{\epsilon_{\hat{\theta}}} (\overline{X}; [X]Y)}.
\end{align}
The average probability of error of the group transmission scheme can be bounded from above by 
% \begin{align*}
\[   \Pr(\textnormal{error}) = \sum_{u\in J}\frac{1}{\abs{J}}  \Pr(E(u)) 
\leq  \epsilon + \sum_{u \in J} \frac{1}{|J|}\sum_{{\hat{\theta}} \neq \bs }  \abs{T_{{\hat{\theta}}}(u)} \exp_2 \bracenv{
 -I_{\textnormal{H}}^{\epsilon_{\hat{\theta}}}(\overline{X}; [X]Y)}.
\]
Exploiting \eqref{eq:T_theta_size}, we get the desired  result in terms of the rate $R$ of the code.
\endproof

\begin{example}\label{ex:Z4Z8_1shot}
Let $J = \Z_4, G = \Z_8$. 
In this example, we have $\mathcal{G}({G}) = \mathset{(2,3,1)}$ and $\mathcal{G}(J) = \mathset{(2,2,1)}$, $k_{2,1} = 0, k_{2,2} = 1, k_{2,3} = 0$, and the term $g_{(2,2,1)\rightarrow (2,3,1)}$  is a uniform random variable over $2\Z_8$. For simplicity, we write $u = u_{2,2,1} \in  J$ and $g = g_{(2,2,1)\rightarrow (2,3,1)} \in \Z_8$. Then $r_2=\max \mathcal{R}_2(G) = 3$ and the set $\mathcal{Q}(J) = \mathcal{S}(G) = \mathset{(2, 1), (2,2), (2,3)}$, and $\bs_{(2,1)} = 1, \bs_{(2,2)} = 2, \bs_{(2,3)} = 3$. 
For distinct $u,\tilde{u}\in J$, the vector $\hat{\theta}=(\hat{\theta}_{2,1}, \hat{\theta}_{2,2}, \hat{\theta}_{2,3})$ for which $\tilde{u} \in  T_{\hat{\theta}}(u)$ must have 
$\hat{\theta}_{2,1} = 1, 0\leq \hat{\theta}_{2,2} <  2, \hat{\theta}_{2,3} =3$. Thus 
$    P_{err}(u) \leq P_{err}(u, (1,0,3)) + P_{err}(u, (1,1,3)).$

The set $\mathcal{Q}(G) = \mathset{(2,3)}$, so $\pmb{\theta}(\hat{\theta}) = \pmb{\theta}(\hat{\theta})_{(2,3)}$ and 
\begin{align*}
\pmb{\theta}(\hat{\theta})_{(2,3)} &=
\min_{\substack{(2,s)\in\mathcal{S}(G)\\w_{2,s}\ne 0}} 
\bracenv{|3-s|^+ +\hat{\theta}_{2,s}} = |3-2|^+ +\hat{\theta}_{2,2} =1+ \hat{\theta}_{2,2}.
\end{align*}

\textit{Case 1: }{$\hat{\theta}_{2,2} = 0$, $\pmb{\theta}(\hat{\theta})_{(2,3)} = 1$}. 
For $\tilde{u} \in  T_{\hat{\theta}}(u)$, 
$\tilde{u} - u \in Z_4 \backslash 2\Z_4$,  
$H_{\hat{\theta}} = 2\Z_8$, and
$|T_{(1,0,3)}(u)|=2$.
Let $A^*_{\epsilon/2, \hat{\theta}}$ be a maximizer for $D_{\textnormal{H}}^{\epsilon/2}(P_{XY} \| P_{[X]_{\hat{\theta}}} P_{X\mid [X]_{\hat{\theta}}} P_{Y\mid [X]_{\hat{\theta}}} )$. 
Thus we have
\begin{align*}
    P_{err}(u, (1,0,3)) \leq \abs{T_{(1,0,3)}(u)}  \exp_2 
\Large\{-I_{\textnormal{H}}^{\epsilon/2}(\overline{X}; [X]Y) \Large\}.
\end{align*}

\textit{Case 2: }{$\hat{\theta}_{2,2} = 1$, $\pmb{\theta}(\hat{\theta})_{(2,3)} = 2$}
For $\tilde{u} \in  T_{\hat{\theta}}(u)$, 
$\tilde{u} - u \in 2Z_4 \backslash 4\Z_4$,  
 $ H_{\hat{\theta}} = 4\Z_8$, 
$|T_{(1,1,3)}(u)|=1$,
and we have 
\begin{align*}
    P_{err}(u, (1,1,3)) \leq  \abs{T_{(1,1,3)}(u)}  \exp_2 
    \Large\{-I_{\textnormal{H}}^{\epsilon_{\hat{\theta}},\hat{\theta}}(\overline{X}; [X]Y) \Large\}.
\end{align*}
Therefore the error probability for a message $u$ is \begin{align*}
 \Pr(E(u)) &\le \Pr(E_1(u))+\Pr(E_2(u)\cap (E_1(u))^c) \\
 &\leq \epsilon + \abs{T_{(1,0,3)}(u)}  \exp_2 \Large\{-I_{\textnormal{H}}^{\epsilon_{\hat{\theta}},\hat{\theta}}(\overline{X}; [X]Y) \Large\}_{\hat{\theta} =(1,0,3)} + 
 \abs{T_{(1,1,3)}(u)}  \exp_2 \Large\{-I_{\textnormal{H}}^{\epsilon_{\hat{\theta}},\hat{\theta}}(\overline{X}; [X]Y) \Large\}_{\hat{\theta} =(1,1,3)},
 \end{align*} where we use $A_{\epsilon} = \cap_{\hat{\theta}} A^*_{\epsilon/2, \hat{\theta} }\,$. 
\end{example}

\subsection{Generalized Achievability}\label{subsub:CC-1Achie_2nd}

In this section, we provide an alternative characterization of an achievable one-shot performance of group codes on classical channels.  
Recall in Section \ref{subsec:H_coset}, for an Abelian group $G$, we defined vectors $\hat{\theta}$ whose components are non-negative integer-valued and indexed by $(p,s)\in\mathcal{S}(G)$, and 
 $\Theta = \Theta(G)$ to be the set of vectors $\hat{\theta}$ with cardinality $M\triangleq \abs{\Theta}.$ 
\begin{definition}\label{def::Htheta_cond_CC_general}
Consider an arbitrary pair $( \eta, b)$ and any vector $\hat{\theta} \in \Theta$.
% , let $[x_r]$ denote $[x_r]= [x_r]_{\hat{\theta}} = x_r + H_{\eta+\hat{\theta}}$. 
Let $P_X, P_{[X]}, P_{X\mid [X]}, P_{Y\mid [X]}$ be defined as in Definition~\ref{def:Htheta_cond_CC}. 
For a representative   $\sparenv{x_r}$ of a coset of $H_{\eta+\hat{\theta}}$ in $H_{\eta}+b$, define two conditional joint distributions $P_{XY \mid [X]}$, $P_{XY\mid \hat{\theta}}$ by
\begin{align}
    P_{XY\mid [X]}(\tilde{x}, y \ \vert \sparenv{x_r}) &\triangleq   P_{X\mid [X]}(\tilde{x} \ \vert \sparenv{x_r}) P_{Y\mid [X]}(y\ \vert  \sparenv{x_r}),  \nonumber \\
     P_{XY\mid \hat{\theta}}(\tilde{x}, y) &\triangleq \sum_{\sparenv{x_r}} P_{[X]}(\sparenv{x_r}) P_{XY\mid [X]}(\tilde{x}, y \ \vert \sparenv{x_r}),  \nonumber 
\end{align} and a joint distribution 
for all $\tilde{x} \in H_{\eta}+b$ and $y \in \mathcal{Y}$:
\begin{align}
    P_{XY\mid J } (\tilde{x}, y) &\triangleq ( \abs{J} -1)^{-1} \sum_{{\hat{\theta}}\in \Theta }  \abs{T_{\hat{\theta}}} P_{XY\mid \hat{\theta}}(\tilde{x}, y) \label{eq:PXYJ_CC-1_2nd} \\
 &= ( \abs{J} -1)^{-1} \sum_{{\hat{\theta}} } \abs{T_{{\hat{\theta}}}} \sum_{\sparenv{x_r}} 
P_{[X]}(\sparenv{x_r})  P_{X\mid [X]}(\tilde{x} \ \vert \sparenv{x_r}) P_{Y\mid [X]}(y\ \vert  \sparenv{x_r}) \nonumber \\
&=( \abs{J} -1)^{-1} \sum_{{\hat{\theta}} }\abs{T_{{\hat{\theta}}}} \frac{{1}}{\abs{H_{\eta}}}
\sum_{{x} \in \sparenv{\tilde{x}}}   \frac{1}{\abs{H_{\eta+\hat{\theta}}}}
W_{Y|X}(y|{x}) .  \label{eq:PXYJ_CC-1_2nd_GH} 
\end{align}
\end{definition}
Note that $P_X(x) = \frac{1}{\abs{H_{\eta}}}$ for all $x \in H_{\eta} +b$, thereby 
\begin{equation}\label{eq:PXYJ_CC-1_semiProd}
    P_{XY\mid J } (\tilde{x}, y) = P_X(\tilde{x})( \abs{J} -1)^{-1} \sum_{{\hat{\theta}} }\abs{T_{{\hat{\theta}}}} \sum_{{x} \in \sparenv{\tilde{x}}}   \frac{1}{\abs{H_{\eta+\hat{\theta}}}} W_{Y|X}(y|{x}).
\end{equation}
We emphasize that the dependence of $P_{XY|J}$ on the pair $(\eta,b)$ is not made explicit till now for notational simplicity. 
Let $ A_{\epsilon}$ be a subset of $\cX\times \cY$ that achieves  $D_{\textnormal{H}}^{\epsilon}(P_{X_{\eta,b}Y} \| P_{X_{\eta,b}Y\mid J })$. That is, 
$P_{X_{\eta,b}Y}(A_{\epsilon}) \geq 1-\epsilon $ and 
\begin{equation}\label{eq:A_epsi_CC-1_2nd}
    D_{\textnormal{H}}^{\epsilon}(P_{X_{\eta,b}Y} \| P_{X_{\eta,b}Y\mid J }) = -\log_2 P_{X_{\eta,b}Y\mid J }(A_{\epsilon}).
\end{equation} 
% where we use $P_{XY\mid J}(A)$ as  a shorthand notation for $\sum_{(x, y) \in A} P_{XY\mid J}({x}, y)$.

%\blue{Interpretation}

We now state the main result of this subsection. 
\begin{theorem}\label{thm:CC1_achieve_2nd}
For a given classical channel $(G,\mathcal{Y},W_{Y|X})$, and any $J$ as given in \eqref{eqn:J},
there exists a $(1,\abs{J} , \epsilon')$-group transmission system  such that 
\[
R\geq \max_{\eta,b} D_{\textnormal{H}}^{\epsilon}(P_{X_{\eta,b}Y} \| P_{X_{\eta,b}Y\mid J }) - 
\log_2 \frac{1}{\epsilon'- \epsilon},\]
for any $\epsilon \in (0,\epsilon')$, where the rate $R = \log_2 \abs{J} $.
\end{theorem}
We have the following proposition which relates the two lower bounds given above,  a proof of which is  given in Appendix \ref{app:prooof_proposition}.
\begin{prop}
    \label{prop:proposition}
    The lower bound on $R$ given in \Tref{thm:CC1_achieve_2nd} is no smaller than that in \Tref{thm:CC1_achieve}, i.e., for any subgroup $J$ as in \eqref{eqn:J} and  $(\eta,b)$ we have  
    \[
  D_{\textnormal{H}}^{\epsilon}(P_{X_{\eta,b}Y} \| P_{X_{\eta,b}Y\mid J }) \geq -\log_2   
\sum_{{\hat{\theta}} \neq \bs }   \exp_2 \bracenv{ 
- (\omega_{\hat{\theta}} R
 -I_{\textnormal{H}}^{\epsilon_{\hat{\theta}}}(\overline{X}_{\eta,b}; [X_{\eta,b}],Y)) }.
    \]
\end{prop}
 \begin{remark}
As noted above, the result of  \Tref{thm:CC1_achieve_2nd} is 
tighter than that of \Tref{thm:CC1_achieve}. However, the latter is 
easier to work with as demonstrated in the application to the 
asymptotic capacity characterization in later sections. 
\end{remark}
\noindent \textbf{Proof of Theorem \ref{thm:CC1_achieve_2nd}:}   
We consider the random coding argument of the proof of Theorem \ref{thm:CC1_achieve} as given in the previous subsection, where the error events $E_1(u)$ and $E_2(u)$ were defined.  
In the following we provide alternative upper bounds on the probabilities of these events.
The probability for the event $ E_1(u)$ is given by
\begin{align}
\mathbb{E} [\Pr(E_1(u))] &= \Pr\parenv{ ( \phi(u) +V , Y)\notin A_{\epsilon} } = 
 1-  \Pr\parenv{ ( \phi(u) +V , Y)\in A_{\epsilon} } \nonumber \\
 &= 1- \sum_{x} \Pr( \phi(u) +V =x)\sum_{y: (x,y)\in A_{\epsilon} } W_{Y\mid X}(y\ \vert x) \nonumber \\
 &= 1- \sum_{(x,y)\in A_{\epsilon} } P_X(x) W_{Y\mid X}(y\ \vert x)
 = 1- P_{XY}(A_{\epsilon}) \leq \epsilon, \label{eq:PrE1u_CC1_2nd}
\end{align} where the expectation and probability are evaluated with the distribution of the code ensemble given in Definition~\ref{def:groupCode}, specifically, the homomorphism $\phi$ and the `shift' $V$.
% $P_X$ is the uniform distribution on $H_{\eta}+b$.

Combining equations \eqref{eq:Perr_utheta_CC-1} and \eqref{eq:Perr_uutilde_CC1_expanded}, 
from Appendix \ref{app:CC_Error} with regard to the analysis of proof of Theorem \ref{thm:CC1_achieve}, we have 
\begin{align}
\mathbb{E}[\mbox{Pr}(E_2(u) \cap E_1(u)^c)]
&\leq  
\sum_{{\hat{\theta}} \neq \bs } \sum_{\tilde{u}\in T_{{\hat{\theta}}}(u)} \sum_{\sparenv{x_r}} \sum_{\tilde{x} \in \sparenv{x_r}} \sum_{y: (\tilde{x}, y)\in A_{\epsilon} } 
P(\sparenv{x_r})  P(\tilde{x} \ \vert \sparenv{x_r}) P(y\ \vert  \sparenv{x_r})  \nonumber  \\
&= \sum_{{\hat{\theta}} \neq \bs } \sum_{\tilde{u}\in T_{{\hat{\theta}}}(u)} \sum_{\sparenv{x_r}} 
\sum_{(\tilde{x}, y)\in A_{\epsilon} } 
P(\sparenv{x_r})  P(\tilde{x}, y \ \vert \sparenv{x_r})  \nonumber 
\end{align}
\begin{align}
&= \sum_{(\tilde{x}, y)\in A_{\epsilon} }  \sum_{{\hat{\theta}} \neq \bs } \sum_{\tilde{u}\in T_{{\hat{\theta}}}(u)}   P_{XY\mid \hat{\theta}}(\tilde{x}, y)  \nonumber \\
&=  
\sum_{(\tilde{x}, y)\in A_{\epsilon} } 
\sum_{{\hat{\theta}} \neq \bs } 
\abs{T_{{\hat{\theta}}}(u)} P_{XY\mid \hat{\theta}}(\tilde{x}, y), \label{eq:Perr_u_full_expand}
\end{align} where the first equality holds because $P(\tilde{x}, y \ \vert \sparenv{x_r}) $ vanishes for $\tilde{x}\notin \sparenv{x_r}$, and the last equality holds because for any $x,y$, the term $P_{XY\mid \hat{\theta}}({x}, y)$ depends only on  $\hat{\theta}$ and is independent of $\tilde{u}$.
Finally, using the definition of $P_{XY|J}$ as in \eqref{eq:PXYJ_CC-1_2nd}, the probability of error for $u$ can be  bounded from above as 
\begin{equation}\label{eq:Perr_u_inJ_CC-1}
\mathbb{E}[\mbox{Pr}(E_2(u) \cap E_1(u)^c)]
  \leq ( \abs{J} -1) \sum_{(\tilde{x}, y)\in A_{\epsilon} } P_{XY\mid J } (\tilde{x}, y) \leq \abs{J} P_{XY\mid J } (A_{\epsilon}).
\end{equation}
We now provide an upper bound for the probability of error for the coding scheme by combining the bounds on the probabilitites of the two error events 
as follows.
\begin{align}
    \Pr(\textnormal{error}) 
&= \frac{1}{\abs{J}} \sum_{u\in J} \Pr(E(u))
\leq \epsilon +  \abs{J} P_{XY\mid J } (A_{\epsilon}) 
= \epsilon + 2^{R - D_{\textnormal{H}}^{\epsilon}(P_{X_{\eta,b}Y} \| P_{X_{\eta,b}Y\mid J })}, 
\end{align} where the inequality follows from  \eqref{eq:PrE1u_CC1_2nd} and  \eqref{eq:Perr_u_inJ_CC-1}. The last equality follows from \eqref{eq:A_epsi_CC-1_2nd} and the rate of the code $R$ being $R = \log_2 \abs{J}$, as in equation~\eqref{eqn:rate}. The characterization of error probability for the coding scheme proves the claim. 
\endproof

\subsection{Converse}\label{subsec:CC1shot_conv}
Toward the converse, we have the following theorem whose proof is given in Appendix \ref{app:CC_Error_C}. This is the main result of this subsection. Given a classical channel 
$(G,\mathcal{Y},W_{Y|X})$, and an input group $J$ as defined in \eqref{eqn:J}, 
and a $(1,|J|,\epsilon)$ group transmission system, 
define the input distribution as 
$P_X$ is uniform on $\{\phi(J)+V\}$, where $\phi$ is the homomorphism and $V$ is the shift. Define 
$P_{[X]},P_{X|[X]}$ and $P_{Y|[X]}$ similar to those in Definition \ref{def:Htheta_cond_CC} for every $\hat{\theta} \neq \bs$ by noting that
$[X]$ is a function of $X$.

\begin{theorem}\label{thm:CC1_conv}
   Given a classical channel  $(G,\mathcal{Y},W_{Y|X})$, and $J$ as given in \eqref{eqn:J}, 
   any $(1, \abs{J}, \epsilon)$ group transmission system satisfies
    \[
    R \leq \min_{\hat{\theta}\neq \bs} \frac{1}{1- \omega_{\hat{\theta}}  } 
    I_{\textnormal{H}}^{\epsilon}(\overline{X}; [X] Y),
    \]
    where the rate $R = \log_2 \abs{J}$.
\end{theorem}

We have provided a converse for the performance characterization as given in Theorem \ref{thm:CC1_achieve}. 

\section{One-shot Classical-Quantum Group Coding}
\label{sec:CQ_1shot}

\subsection{Achievability}\label{subsec:CQ_1shot_achieve}
Consider an arbitrary pair of output and input Abelian groups $G$ and $J$, a CQ channel $\{\rho_x\}_{x \in G}$, 
a pair $(\eta,b)$
and a subgroup $H_{\eta+\hat{\theta}}$ of $G$ indexed by $\hat{\theta}$. We assume the uniform distribution on $H_{\eta}+b$ for 
the input of the CQ channel yielding the input-output joint state as 
\begin{equation}\label{eq:CQ_joint_rhoAB}
\rho^{AB} \triangleq \sum_{x \in H_{\eta}+b} \frac{1}{|H_{\eta}|} |x\ket\bra x|^A \otimes \rho^B_x,    
\end{equation}
where $\mathcal{X}$ denotes\footnote{Recall that the  classical variable (channel input) $X$ with alphabet $\mathcal{X}$ is stored in a quantum register with Hilbert space also denoted as $\mathcal{X}.$}   the input space and $\mathcal{B}$ denotes the output space. We refer to the input quantum state as $A$ and the output quantum state as $B$. 
Recall that for an element $x \in H_{\eta}+b$ there is a one-to-one mapping $x \leftrightarrow ([x]_{\hat{\theta}}, \overline{x}_{\hat{\theta}})$, where $[x]_{\hat{\theta}}$ is the representative of the coset of $H_{\eta+\hat{\theta}}$ which $x$ belongs to, and $\overline{x}_{\hat{\theta}} \in H_{\eta+\hat{\theta}}$, such that $x =[x]_{\hat{\theta}} + \overline{x}_{\hat{\theta}}$.  
Define the transversal, the set of coset representatives of $H_{\eta+\hat{\theta}}$ in $G$,  as $T^G_{\hat{\theta}}$. 
We drop $\hat{\theta}$ from the subscript, when it is clear from the context.

\begin{definition}
\label{def:CQ_H_theta}
For any $H_{\hat{\theta}}$, 
using the one-to-one mapping $x \leftrightarrow ([x]_{\hat{\theta}}, \overline{x}_{\hat{\theta}})$,
the joint state can be viewed as follows
\begin{align*}
\rho^{AB}&=    \rho^{[A] \bar{A} B} \triangleq \sum_{[x], \overline{x}} \frac{1}{\abs{G}} 
  |[x], \overline{x}\ket\bra [x], \overline{x}|^{[A] \bar{A}}\otimes\rho_{[x], \overline{x}}^B,
  &=  \sum_{[x], \overline{x}} P_{[X]}([x]) P_{\overline{X}}(\overline{x})
  |[x] \ket\bra [x] |^{[A] } \otimes
   | \overline{x}\ket\bra   \overline{x}|^{ \bar{A}}
  \otimes\rho_{[x], \overline{x}}^B \ , 
\end{align*}
and we let $ P_{[X]}([x]) = \frac{\abs{H_{\eta+\hat{\theta}}}}{\abs{H_{\eta}}}$ for all $[x]$ and $ P_{\overline{X}}(\overline{x}) = \frac{1}{\abs{H_{\eta+\hat{\theta}}}}$ for all $\overline{x}$. Here $[A]$ denotes the state associated with $T^G_{\hat{\theta}}$ and $\bar{A}$ as that associated with $H_{\eta+\hat{\theta}}$. 
Define 
\begin{align*}
    &\rho^{B}_{[x]} \triangleq \sum_{\overline{x}} P_{\overline{X}}(\overline{x})\rho_{[x], \overline{x}}^B\ ,  \ \ 
     \rho^{\bar{A}} \triangleq \sum_{ \overline{x}} P_{\overline{X}}(\overline{x})  | \overline{x}\ket\bra   \overline{x}|^{ \bar{A}},  \\ 
   &\rho^{[A] B} \triangleq \sum_{[x]} P_{[X]}([x]) |[x] \ket\bra [x] |^{[A] } \otimes \rho^{B}_{[x]} \ .  
\end{align*}
Also define the hypothesis testing mutual information:
\begin{align*}
    I_{\textnormal{H}}^{\epsilon}(\overline{X}_{\eta,b}; [X_{\eta,b}],Y) &\triangleq D_{\textnormal{H}}^{\epsilon}(\rho^{[A] \bar{A} B} || \rho^{\bar{A} } \rho^{[A] B} ) \ .
\end{align*}
\end{definition}
For simplicity, we use the same 
notation $I_{\textnormal{H}}^{\epsilon}$ to denote both classical as well as quantum one-shot hypothesis testing mutual information. The particular form used should be clear from the context. 
With these definitions, and recalling the quantities $\zeta(G)$ and $\Theta(G)$ (see Section \ref{section:Abelian}) associated with $G$, we are ready to state the first main result of this section. 
\begin{theorem}\label{thm:CQ1_achieve_1}
Let $\epsilon$ and $\mathset{\epsilon_{\hat{\theta}}} $ be given with $ \epsilon_{\hat{\theta}}>0$ for all $\hat{\theta}$ and  $\sum_{\hat{\theta} \neq \bs}  \epsilon_{\hat{\theta}} \leq \epsilon$. 
For a given CQ channel $\mathcal{N}=\{\rho_x\}_{x \in G}$, any $(\eta,b)$, 
and  $J$ as given in \eqref{eqn:J},
 there exists a $(1,J,\epsilon')$-group transmission system such that 
\[
\epsilon' \leq  \sqrt{\epsilon}[33+16|\Theta| \zeta(G)] + 8\sum_{{\hat{\theta}} \neq \bs }   \exp_2 \bracenv{ 
(1- \omega_{\hat{\theta}} )R
 -I_{\textnormal{H}}^{\epsilon_{\hat{\theta}}}(\overline{X}_{\eta,b}; [X]_{\eta,b}, Y) },\]
where the rate $R$ is given in \eqref{eqn:rate}.
\end{theorem}
\begin{remark}
The characterization for the classical-quantum channel mirrors that for the classical channel. The similarity ends there. The proof is significantly more involved in the quantum case because of the following reasons. The intersection of decision regions used in the proof of the classical case cannot be implemented simply as intersection of operators. We have to use a synergy of complex machinery of intersection of projectors as given in \cite{sen2021unions} and the intricate group coding approach developed in the classical setting toward developing  a proof. 
Although this is a point-to-point channel coding problem, the 
lattice of subgroups of $J$ impose a host of constraints on the communication. Any solution to the problem has  similar issues to address as those 
in classical-quantum multiple-access channels with an arbitrary number of transmitters.  In fact, it is compounded by the complex structure of dependency among $[u],\overline{u}$, $[x]$ and $\overline{x}$ for every  subgroup $H_{\hat{\theta}}$ of input group $J$ as delineated in Remark \ref{rem:dependency}.
\end{remark}

\begin{remark}
     The theorem motivates the definition of one-shot $\epsilon$-group capacity as
    \[
    \max_{\eta,b} \min_{\hat{\theta} \neq \mathbf{s}}  \frac{1}{(1-w_{\hat{\theta}})} 
    I_{\textnormal{H}}^{\epsilon_{\hat{\theta}}}(\overline{X}_{\eta,b}; [X_{\eta,b}],Y).
    \]
    We compute this quantity for several examples in Section \ref{sec:examples}.
\end{remark}

\noindent \textbf{Proof:}
We give a detailed proof for the case when $\eta=\zero$, and $b$
 is arbitrary. 
 An extension to any general $\eta$ vector is straightforward by redefining the group $G$ accordingly as $H_{\eta}$, and by redefining the channel as first adding $b$ to the input and then 
 acting with random transformation given by $\mathcal{N}$. 
 
Let a set of parameters $\mathset{\epsilon_{\hat{\theta}}}_{{\hat{\theta}} \neq \bs}$ be given such that  ${\epsilon_{\hat{\theta}}} >0$ for each $\hat{\theta}$ and that $\sum_{\hat{\theta}} \epsilon_{\hat{\theta}} = \epsilon$. 
Consider the optimizing POVM $(\Pi'')_{[X]} $ in $[A]\bar{A}B$ arising in the definition of $ I_{\textnormal{H}}^{\epsilon_{\hat{\theta}}}(\overline{X}; [X]Y)$.
The POVM satisifies:
\begin{align}
    (\Pi'')_{[X]} &= \sum_{[x], \overline{x}} 
    |[x], \overline{x}\ket\bra [x], \overline{x}|^{[A]\bar{A}} 
    (\Pi'')^B_{[X]; [x], \overline{x}}, \nonumber \\
    \label{eq:cq_nshot_POVM_mutual_new}
    \trace\sparenv{(\Pi'') \rho^{A B}} &\geq 1- \epsilon_{\hat{\theta}}\ , \quad 
    \trace\sparenv{(\Pi'') \rho^{\bar{A}}\otimes  \rho^{[A] B}} \leq 2^{-  I_{\textnormal{H}}^{\epsilon_{\hat{\theta}}}(\overline{X}; [X]Y)}.
    \end{align}
We consider an extended version of the CQ channel $\mathcal{N}$ as follows. 
Define $ \widehat{\mathcal{B}} \triangleq \mathcal{B} \otimes \CC^2$. 
By Fact 2 in \cite{sen2021unions}, there are orthogonal projections $ \Pi^{\widehat{B}}_{[X];x}$ in $ \widehat{\mathcal{B}} $ that give the same measurement probability on states $\sigma^B \otimes |0 \ket\bra 0 |$ that POVM elements 
$(\Pi'')^B_{[X]; x}$ give on states $\sigma^B$. 
Let $W_{[X];x}$ denote the orthogonal complement of the support of $ \Pi^{\widehat{B}}_{[X];x}$ in $\widehat{\mathcal{B}}$. 

\noindent \textbf{Step 1:}  Consider a new Hilbert space $\mathcal{L}$ that is used only as a quantum register to store classical values, and  
define the extended output space 
\[
 \mathcal{B}' \triangleq  (\mathcal{B}\otimes \CC^2) \oplus 
 \bigoplus_{(p,r,m,k) \in \mathcal{G}^*(G)}
  (\mathcal{B}\otimes \CC^2 \otimes \mathcal{L}^{(p,r,m,k)}).
 \]
where 
 $\mathcal{L}^{(p,r,m,k)}$ is isomorphic to $\mathcal{L}$.

\begin{remark}
Recall 
$\zeta(G)= \sum_{(p,r,m) \in \mathcal{G}(G)} r$, 
the sum of prime powers in the prime factorization of $|G|$. 
Also recall  the set $\mathcal{G}^*(G) \triangleq \{(p,r,m,k): (p,r,m) \in \mathcal{G}(G), 1 \leq k \leq r\}$.
We use a vector representation for the elements of $G$ as follows. First we note that
every element of $a$ of $\mathbb{Z}_{p^r}$ can be represented uniquely as a $r$-length vector $(a_1,a_2,\ldots,a_r)$   such that 
$a_i$ takes value in the transversal of   $p^{r-i}\mathbb{Z}_{p^r} / p^{r-i+1}\mathbb{Z}_{p^r}$.
A shorter vector $(a_1,a_2,\ldots,a_k)$ can 
represent any element in the subgroup $p^{r-k} \mathbb{Z}_{p^r}$. 
This is extended to all of $G$ using the direct sum operation. We denote this mapping as $\nu(a)$. 
Hence every element of $G$ is represented 
uniquely as a vector of length $\zeta(G)$. This is also extended to all subgroups of $G$. 
Furthermore, we use 
$(p,r,m,k) \in \mathcal{G}^*(G)$ to  index the elements of a vector. The set formed by the   $(p,r,m,k)$-th
elements of $\nu(G)$ is denoted as $\mathcal{X}_{(p,r,m,k)}$. This also denotes the Hilbert space associated with the quantum register that stores the corresponding classical values, where a direct sum is replaced with a tensor product.
\end{remark}

We extend the space corresponding to the alphabet of $(p,r,m,k)$-th element of $\nu(G)$ as follows. 
For every 
$(p,r,m,k) \in \mathcal{G}^*(G)$  define 
\[
\mathcal{X}'_{(p,r,m,k)}=\mathcal{X}_{(p,r,m,k)} \otimes \mathcal{L}^{(p,r,m,k)}.
\]
This leads to the following extensions. For each subgroup $p^{r-s}\mathbb{Z}_{p^r}^{(m)}$, $s=1,2,\ldots,r$, the extended space is 
\[
p^{r-s}\mathbb{Z}_{p^r}^{(m)}
\otimes \bigotimes_{k=1}^s \mathcal{L}^{(p,r,m,k)},
\]
and for the transversal of  
$\mathbb{Z}_{p^r}^{(m)} / p^{r-s}\mathbb{Z}_{p^r}^{(m)}$, 
the extended input space  is given by
\[
(\mathbb{Z}_{p^r}^{(m)} / p^{r-s}\mathbb{Z}_{p^r}^{(m)})
\otimes \bigotimes_{k=s+1}^r \mathcal{L}^{(p,r,m,k)}.
\]

\noindent \textbf{Step 2:} Let $0 < \delta < 1$.
Consider a vector $\Bell $ indexed by 
$(p,r,m,k)$ for $(p,r,m,k) \in \mathcal{G}^*(G)$, where  $\ell_{(p,r,m,k)} \in \mathcal{L}^{(p,r,m,k)}$, and is a basis element of $\mathcal{L}^{(p,r,m,k)}$.
Let $\bar{\Bell}_{\hat{\theta}}$ denote the sub-vector corresponding to the subgroup $H_{\hat{\theta}}$, 
and the complementary sub-vector $[\Bell]_{\hat{\theta}}$ corresponding to the transversal $T_{\hat{\theta}}$.  For the trivial case
$H_{\hat{\theta}}=\{0\}$, we have $[\Bell]=\Bell$. Recall that in this case, $[X]=X$. 

For any  subgroup $H_{\hat{\theta}}$ of $G$, 
define the \textit{tilting} map $\mathcal{T}_{[X]; [\Bell], \delta}:  \widehat{\mathcal{B}} \goto   \mathcal{B}' $ defined as 
\begin{align*}
    \mathcal{T}_{[X]; [\Bell], \delta}: |h\ket &\mapsto \frac{1}{\sqrt{1+\zeta(G)\delta^2}}\big(|h\ket  
     +\sum_{(p,r,m,k) \in \mathcal{G}^*(G) / \mathcal{G}^*(H)} \delta |h\ket | {l_{(p,r,m,k)}} \ket \big)\ .
\end{align*}
Note that we are tilting only along the direction of transversal of $H_{\hat{\theta}}$ in $G$. 
% Define the tilting map $\mathcal{T}_{[x]; l, \delta}$  analogously.
Define a state 
\begin{equation*}
    ({\rho'})^{  B'}_{x,\Bell,\delta} \triangleq 
    \mathcal{T}_{X; \Bell, \delta}\parenv{
        \rho_{x}^B \otimes | 0 \ket\bra 0 |^{\CC^2 }
    }.
\end{equation*}
Consider the classical–quantum state 
\begin{equation}
    ({\rho'})^{A' B'} \triangleq \abs{\mathcal{L}}^{-\zeta(G)} \sum_{x,\Bell } 
    P_{X}(x) 
    |x,\Bell\ket\bra x,\Bell|^{A' } 
  \otimes 
  ({\rho'})^{  B'}_{x,\Bell, \delta}\ .
\end{equation}
This corresponds to the extended channel $\mathcal{N}'$ with inputs $x$ and $\Bell$, and output state
 $({\rho'})^{  B'}_{x,\Bell, \delta}$
in $\mathcal{B}'$. We will work with this channel instead of the given one $\mathcal{N}$,  until the very end. 
It can be shown that
\begin{equation}
\label{eq:closeness_rho_new}
\left \|  
  ({\rho'})^{  B'}_{x,\Bell,\delta} -
        \rho_{x}^B \otimes | 0 \ket\bra 0 |^{\CC^2}  
\right\|_1 \leq  2\zeta(G) \delta^2.  
\end{equation}
Define the tilted space
\begin{align*}
    W'_{[X]; x,\Bell, \delta} &\triangleq \mathcal{T}_{[X]; [\Bell], \delta}(W_{[X];x})
\end{align*} residing in $\mathcal{B}'$. 
Define the subspace 
\begin{equation}
\label{eq:sum_projection_new}
    W'_{x,\Bell,\delta} \triangleq 
   \bigoplus_{\hat{\theta} \in \Theta} W'_{[X]; x, \Bell,  \delta} ,
\end{equation} and 
$  (\Pi')^{B'}_{W'} = (\Pi')^{B'}_{W'_{x,\Bell, \delta}} $ the orthogonal projection in $\mathcal{B}'$ onto $W'_{x,\Bell,\delta}$. 
Let $\Pi^{B'}_{\widehat{B}} $ be the orthogonal projection in $\mathcal{B}'$ onto $\widehat{\mathcal{B}}$. Define POVM elements 
\begin{equation}
\label{eq:final_POVM_new}
      (\Pi')^{B'}_{x,\Bell, \delta} \triangleq 
      \parenv{ \one^{B'} - (\Pi')^{B'}_{W'}} \Pi^{B'}_{\widehat{B}} \parenv{ \one^{B'} - (\Pi')^{B'}_{W'}}\ ,
\end{equation}
\begin{equation}
    (\Pi')^{A' B'} \triangleq \sum_{x,\Bell}
    |x,\Bell \ket\bra x,\Bell |^{A' } 
  \otimes (\Pi')^{B'}_{{x,\Bell, \delta}}\ .
\end{equation}
Define the following states for any subgroup $H_{\hat{\theta}}$:
\[
(\rho')^{B'}_{[x],[\Bell]} \triangleq 
\frac{1}{|\mathcal{L}|^{\zeta(H)}} \sum_{\overline{x},\bar{\Bell}} P(\overline{x}) 
(\rho')^{B'}_{x,\Bell},
\]
\[
(\rho')^{B'} \triangleq 
\frac{1}{|\mathcal{L}|^{\zeta(G)}} \sum_{x,\Bell} P(x)  
(\rho')^{B'}_{x,\Bell}.
\]
We make the following observations using the arguments given in \cite{sen2021unions}:
\begin{align*}
(\rho')^{B'}_{[x],[\Bell]} &= \frac{1+(\zeta(G)-\zeta(H)))\delta^2}{1+\zeta(G) \delta^2} \mathcal{T}_{[X];[\Bell]}(\rho_{[x]}^B \otimes |0\ket\bra 0|) 
+ N_{[X];[x],[\Bell]},
\end{align*}
\[
(\rho')^{B'} = \frac{1}{1+\zeta(G) \delta^2}  (\rho^B \otimes |0\ket\bra 0|)+ N,
\]
for some operators satisfying 
\begin{equation}
\label{eq:l_infinity_bound_new}
\| N_{[X];[x],[\Bell]} \|_{\infty} \leq \frac{2\zeta(G) \delta}{\sqrt{|\mathcal{L}|}}, 
\quad
\| N \|_{\infty} \leq \frac{2 \zeta(G) \delta}{\sqrt{|\mathcal{L}|}}.
\end{equation}
Furthermore, using Holder's inequality,
we have 
\begin{equation}
\label{eq:l_1_bound_new}
\| (\Pi')^{B'}_{x,\Bell} \|_1 \leq 2 |\mathcal{B}|.
\end{equation}
\noindent \textbf{Step 3:} Probability of Error Analysis:
We construct a random code as follows. We generate the random homomorphism as stated in the previous section that maps 
$u\in J$ to $x \in G$ according to 
$x=\phi(u)+V$. 
In addition, we generate for every $u \in J$, a random vector $\Bell$ independently and uniformly such that 
$[\Bell]$ depends only on $[u]$ and $\bar{\Bell}$ depends on the entire $u$. 
These together determine the random encoding function $u \mapsto (x,\Bell)(u)$.
The decoding POVM is constructed from
the POVM elements 
$(\Pi')^{B'}_{x,\Bell, \delta}$.  
We use the square root measurement \cite{holevo1998capacity,schumacher1997sending} from the collection of operators:
$\{(\Pi')^{B'}_{(x,\Bell)(u), \delta}\}_{u \in J}$.

We start by computing the average probability of error on the extended channel $\mathcal{N}'$ for a fixed $u$ using Hayashi-Nagaoka inequality \cite{hayashi2003general} as follows:
\begin{align}
   \EE_{C}&(P(\mbox{Error}|u)) = \EE_{C} \trace\sparenv{ \parenv{
    \one^{B'} - \Lambda^{B'}_{u} }
    ({\rho'})^{B'}_{u, \delta }
 } \nonumber \\
\leq &2 \EE_{C} 
\trace\sparenv{ \parenv{
    \one^{B'} - (\Pi')^{B'}_{(x, \Bell)(u) }({\rho'})^{B'}_{(x, \Bell)(u), \delta} }
 }
+ 4 \sum_{\hat{\theta} \in \Theta} \sum_{u'\in T_{\hat{\theta}}(u) } \EE_{C} \trace \bigg[ (\Pi')^{B'}_{([x], [\Bell])( [u']), (\overline{x},\bar{\Bell}) (u')}  ({\rho'})^{B'}_{([x], [\Bell)( [u]), (\overline{x},\bar{\Bell}) (u), \delta} \bigg].
\label{eq:CQ_achieve_simplify}
\end{align} 
We show in  Appendix \ref{app:CQ_Error} that  upper bound on the expected probability of error for a given $u$ can be simplified as 
\begin{align*}
   \EE_{C}(P(\mbox{Error}|u)) \leq 
   32\zeta(G)\delta^2 +\frac{8(1+\zeta(G)\delta^2) |\Theta| \epsilon}{\delta^2} 
   + 8 \sum_{\hat{\theta} \in \Theta} |T_{\hat{\theta}}| 2^{-I_{\textnormal{H}}^{\epsilon_{\hat{\theta}}}(\overline{X},[X];B)},
   \end{align*}
for large enough $|\mathcal{L}|$.
Since the right hand side is independent of $u$, the same bound applies to expected probability of error for the CQ channel. 
Using the bound given in (\ref{eq:closeness_rho_new}), it follows that the expected average probability of error for the given channel $\mathcal{N}$ is at most $\zeta(G) \delta^2$ added to the above upper bound. Thus there exists a group coding scheme whose average probability of error satisfies the bound given in the theorem by 
choosing
$\zeta(G) \delta^2=\sqrt{\epsilon}$.
This completes the proof. 
\endproof
 
 % The details are given in Appendix \ref{app:CQ_Error}. 

\subsection{Generalized Achievability} \label{subsec:CQ1_achie_2nd}
Our next objective is to derive a stricter and a more compact lower bound on the rate than what is given in Theorem \ref{thm:CQ1_achieve_1}.
Consider Abelian groups $G$ and $J$ as in Section~\ref{subsec:AbelianGroup}. 
Given a classical-quantum channel $\cqN = \mathset{\rho_x}_{{x\in \cX}} $ from the classical alphabet $\mathcal{X}$ to the space $\mathcal{B}$, where $\mathcal{X}=G$ is an Abelian group, 
let $u$, $x$ and $\rho_x$ be the message, the channel input and the channel output respectively.
Consider the group code $\mathds{C}=\{\phi(a)+V|a\in J\},$ where the distributions of $\phi$ and $V$ are specified in Definition~\ref{def:groupCode}.
Consider an arbitrary pair $(\eta,b)$.
Let $\pi^{{AB}}$ denote the joint state of the input and output for an input chosen according to the uniform distribution $P_X\equiv \textnormal{Unif}(H_{\eta}+b)$, i.e.,
\begin{align*}
  \pi^{{AB}}  \triangleq  \sum_{x\in H_{\eta}+b } P_X(x)
  |x\ket\bra x|^A\otimes\rho_x^B \ ,
\end{align*}
for any representation of the inputs $x$ in terms of orthonormal vectors $|x \ket^A$ on a Hilbert space $\mathcal{X}$, and where $\pi^{A}$ and $\pi^{B}$ are the corresponding marginals states.
Let $\hat{\theta}$ be a vector indexed by $(p,s)\in \mathcal{S}(G)$ with $0\leq \hat{\theta}_{p,s}\leq s$.
Define the joint state $\pi^{{AB}}_{\hat{\theta}}$ by 
\begin{align*}
  \pi^{{AB}}_{\hat{\theta}}  &\triangleq  \sum_{x\in H_{\eta}+b} P_X(x)\sum_{ \tilde{x}\in x+H_{\eta+\hat{\theta}}} P_{X\mid [X]}(\tilde{x} | x+H_{\eta+\hat{\theta}})
  | \tilde{x}\ket\bra  \tilde{x}|^A\otimes\rho_{ {x}}^B\\ 
  &= \sum_{\sparenv{x_r}}P_{[X]}(\sparenv{x_r}) 
\sum_{{x} \in \sparenv{x_r}}
\sum_{\tilde{x} \in \sparenv{x_r}}
 P_{X\mid [X]}({x} |[x_r]) P_{X\mid [X]}(\tilde{x} |[x])
 | \tilde{x}\ket\bra  \tilde{x}|^A\otimes\rho_{ {x}}^B \\
&= \sum_{\sparenv{x_r}}P_{[X]}(\sparenv{x_r}) 
\sum_{ \tilde{x} \in \sparenv{x_r}}
 P_{X\mid [X]}({x} |[x_r]) 
 | \tilde{x}\ket\bra  \tilde{x}|^A\otimes\rho_{ [x_r]}^B \ ,
\end{align*}
where ${\sparenv{x_r}} ={\sparenv{x_r}}_{\hat{\theta}}$ is the coset $x_r + H_{\eta+\hat{\theta}}$. 
Define a joint state  $\pi^{{AB}}_{J}$ by 
\begin{align}\label{eq:Pi_J_def}
    \pi^{{AB}}_{J} \triangleq( \abs{J} -1)^{-1} \sum_{\hat{\theta} \in \Theta} \abs{T_{\hat{\theta}}} 
     \pi^{{AB}}_{\hat{\theta}} =  \sum_{\hat{\theta} \in \Theta} 2^{\parenv{1- \omega_{\hat{\theta}}} } \pi^{{AB}}_{\hat{\theta}}.
\end{align} The last equality holds by noting that $\log \abs{T_{\hat{\theta}}(a)} = (1- \omega_{\hat{\theta}} ) R$ according to \eqref{eq:T_theta_size}, and that the rate of the code is $R= \log |J|.$
One may also plug in the definition of $ \pi^{{AB}}_{\hat{\theta}}$ and show that 
\begin{align}
    \pi^{{AB}}_{J} &= ( \abs{J} -1)^{-1} \sum_{\hat{\theta} \in \Theta} \abs{T_{\hat{\theta}}} 
\sum_{x\in H_{\eta} +b} P_X(x)\sum_{ \tilde{x}\in x+H_{\eta+\hat{\theta}}} P_{X\mid [X_r]}(\tilde{x} | x+H_{\eta+\hat{\theta}})
| \tilde{x}\ket\bra  \tilde{x}|^A\otimes\rho_{ {x}}^B \nonumber 
\\&= 
( \abs{J} -1)^{-1} \sum_{{\hat{\theta}} } \sum_{\tilde{u}\in T_{{\hat{\theta}}}(u)} \sum_{x , \tilde{x}}  
\Pr\left(\phi(u)+V=x, \phi(\tilde{u})+V=\tilde{x} \right)
| \tilde{x}\ket\bra  \tilde{x}|^A\otimes\rho_{ {x}}^B, \label{eq:Pi_J_exp}
\end{align} where $u$ is an arbitrary element of $J$.

Let $Q^{\ast}$ be a positive operator that achieves\footnote{Now we are making the dependence of the state on the pair $(\eta,b)$ explicit.} $\DHy^{\epsilon }(\pi_{\eta,b}^{{AB}}  \| \pi_{\eta,b, J}^{{AB}} )$ for any $1>\epsilon>0$. That is,  $\trace[Q^{\ast}  \pi_{\eta,b}^{A B}]\geq 1 -\epsilon $ and 
\begin{align} 
\DHy^{\epsilon}(\pi_{\eta,b}^{{A B}}  \| \pi^{{A B}}_{\eta,b, J} ) = - \log_2  \trace[Q^{\ast}  \pi^{{AB}}_{\eta,b, J} ]. \nonumber
\end{align}

We define the decoding POVM by its elements,
\begin{align*}
E_u = \left(\sum_{u'\in J} A_{x(u')} \right)^{-\frac{1}{2}}
A_{x(u)}
\left(\sum_{u'\in J} A_{x(u')} \right)^{-\frac{1}{2}},
\end{align*}
where $x(u) \triangleq \phi(u)+V$ and  $A_x \triangleq \textnormal{tr}_{A }\left[
  \left(|x\ket\bra x|^A \otimes I^B \right) Q^{\ast} \right]$. 
For any state $\rho^B$ on the Hilbert space $\mathcal{B}$, we have 
\begin{align}
    \trace\bracenv{A_x \rho^B}  = \trace\bracenv{\left(|x\ket\bra x|^A \otimes \rho^B  \right) Q^{\ast}}.\label{eq:partial_Q_Ax_CQ-1}
\end{align}

For a specific choice of  $\phi$ and $V$ and a message $u\in J$, the probability of error is given by
\begin{align*}
\Pr(\textnormal{error}|u, \phi, V) = \trace[{(I-E_u)\rho_{x(u)}}].
\end{align*}
Applying Hayashi-Nagaoka inequality \cite{hayashi2003general} with $S = A_{x(u)}$ and $T= \sum_{u' \neq u} A_{x(u')}$, we may bound this term by
\begin{align*}
\Pr(\textnormal{error}|u, \phi, V)  \le 
(1+c)\left(1-\trace[A_{x(u)}\rho_{x(u)}]\right)
+ (2+c+c^{-1})\sum_{u' \neq u}
\trace[A_{x(u')} \rho_{x(u)}],
\end{align*}
for all $c>0.$
Taking expectation over all choices of $\phi, V$, but keeping the transmitted message $u$ fixed, we find
\begin{align}\label{eq:Pe_cq_bound_1-shot}
 \Pr(\textnormal{error}|u) \le (1+c)
 \left[1- \mE_{\phi, V}\ \trace[A_{X(u)}\rho_{X(u)}]\right]
 +   (2+c+c^{-1})\sum_{u' \neq u}
\mE_{\phi, V}\trace[A_{X(u')} \rho_{X(u)}].
\end{align}
Rewriting the first expectation via \eqref{eq:partial_Q_Ax_CQ-1} and the linearity of the trace operator, 
\begin{align}
    \mE_{\phi, V}\trace[A_{X(u)}\rho_{X(u)}]
&= \mE_{\phi, V}\trace\sparenv{ \left(|X(u)\ket\bra X(u)|^A \otimes \rho_{X(u)}^B  \right)Q^{\ast}} \nonumber  \\
&=\trace\sparenv{  \mE_{\phi, V}\left(|X(u)\ket\bra X(u)|^A \otimes \rho_{X(u)}^B  \right) Q^{\ast}} \nonumber \\
&=\trace\sparenv{ Q^{\ast} \pi^{{AB}}  }\geq 1- \epsilon  \nonumber.
\end{align}
Hence we may bound the first term by 
\[
(1+c)
\left[1- \mE_{\phi, V}\ \trace[A_{X(u)}\rho_{X(u)}]\right] \leq (1+c) \epsilon \; .
\]

For the sum in the second term of \eqref{eq:Pe_cq_bound_1-shot},  we adopt an approach similar to that in Section~\ref{subsec:CC1shot_achieve}.
\begin{align}
    \sum_{u' \neq u}
\mE_{\phi, V}\trace[A_{X(u')} \rho_{X(u)}]
&= \sum_{{\hat{\theta}} \neq \bs }
\sum_{
\tilde{u}\in T_{{\hat{\theta}}}(u)}
\sum_{x, \tilde{x} } 
\Pr\left(\phi(u)+V=x, \phi(\tilde{u})+V=\tilde{x}
\right)
\trace[A_{\tilde{x}} \rho_{x}] \nonumber \\
&= \sum_{{\hat{\theta}} \neq \bs }
\sum_{
\tilde{u}\in T_{{\hat{\theta}}}(u)}
\sum_{\sparenv{x_r}}
\sum_{x \in \sparenv{x_r}} 
\sum_{\tilde{x} \in \sparenv{x_r}}
\Pr\left(\phi(u)+V=x, \phi(\tilde{u})+V=\tilde{x}
\right)
\trace[A_{\tilde{x}} \rho_{x}] \nonumber \\
&= \sum_{{\hat{\theta}} \neq \bs } 
P_{err}(u, \hat{\theta}), \label{eq:Pe_theta_cq1}
\end{align} where
$    P_{err}(u, \hat{\theta}) \triangleq \sum_{\tilde{u}\in T_{{\hat{\theta}}}(u)} P_{err}(u,\tilde{u})$ and, for $\tilde{u} \in  T_{{\hat{\theta}}}(u)$, 
\begin{align}
    P_{err}(u,\tilde{u}) &\triangleq 
\sum_{\sparenv{x_r}}
\sum_{x \in \sparenv{x_r}} 
\sum_{\tilde{x} \in \sparenv{x_r}}
\Pr\left(\phi(u)+V=x, \phi(\tilde{u})+V=\tilde{x}
\right)
\trace\left[A_{\tilde{x}} \rho_{x}\right] \nonumber \\
&= 
\sum_{\sparenv{x_r}}P_{[X]}(\sparenv{x_r}) 
\sum_{x \in \sparenv{x_r}} 
\sum_{\tilde{x} \in \sparenv{x_r}}
P_{X\mid [X]}({x} |[x_r])
P_{X\mid [X]}(\tilde{x} |[x_r])
\trace\left[A_{\tilde{x}} \rho_{x}\right] \nonumber \\
&= 
\sum_{\sparenv{x_r}}P_{[X]}(\sparenv{x_r}) 
\sum_{\tilde{x} \in \sparenv{x_r}}
P_{X\mid [X]}(\tilde{x} |[x_r])
\trace\left[A_{\tilde{x}} \rho_{\sparenv{x_r}}\right].
\end{align} 
Leveraging equation~\eqref{eq:partial_Q_Ax_CQ-1}, the above equation can be written as 
\begin{align*}
        P_{err}(u,\tilde{u}) 
=\sum_{\sparenv{x_r}}P_{[X]}(\sparenv{x_r}) 
\sum_{\tilde{x} \in \sparenv{x_r}}
P_{X\mid [X]}(\tilde{x} |[x_r])
\trace\left[\left(|\tilde{x}\ket\bra \tilde{x}|^A \otimes \rho_{\sparenv{x_r}} \right) Q^{\ast} \right]= \trace[Q^{\ast} \pi^{{AB}}_{\hat{\theta}} ].
\end{align*} 
Therefore the sum in \eqref{eq:Pe_theta_cq1} is given by 
\begin{align}
\sum_{u' \neq u} \mE_{\phi, V}\trace[A_{X(u')} \rho_{X(u)}]
&= \sum_{{\hat{\theta}} \neq \bs } \abs{ T_{{\hat{\theta}}}(u)} \trace[Q^{\ast} \pi^{{AB}}_{\hat{\theta}} ] 
% \leq  \sum_{{\hat{\theta}} \in \Theta^{\ast}} \abs{ T_{{\hat{\theta}}}(u)} \trace[Q^{\ast} \pi^{{AB}}_{\hat{\theta}} ] 
\nonumber \\
&= \trace\left[Q^{\ast} \parenv{ \sum_{{\hat{\theta}} } \abs{ T_{{\hat{\theta}}}(u)} \pi^{{AB}}_{\hat{\theta}}} \right]
= \trace\left[Q^{\ast} \parenv{ \sum_{{\hat{\theta}} } 2^{(1- \omega_{\hat{\theta}} )} 2^R \pi^{{AB}}_{\hat{\theta}}} \right] \nonumber \\
&= 2^R \, \trace[Q^{\ast}  \pi^{{AB}}_{J} ] 
= 2^{{R - \DHy^{\epsilon  }(\pi^{{AB}}  \| \pi^{{AB}}_{J} ) }}. \nonumber
\end{align}
The bound in \eqref{eq:Pe_cq_bound_1-shot} reduces to 
\begin{align*}
 \Pr(\textnormal{error}|u) \le (1+c) \epsilon 
 +   (2+c+c^{-1})2^{{R - \DHy^{\epsilon  }(\pi^{{AB}}  \| \pi^{{AB}}_{J} ) }}, 
\end{align*} which is independent of $u$. Therefore we have demonstrated the existence of a $(1,2^R,\epsilon^{\prime})$-code with
\begin{align}\label{eq:lower_cq-1}
  R \ge  \DHy^{\epsilon }(\pi^{{AB}}  \| \pi^{{AB}}_{J} ) 
  -\log_2\frac{2+c+c^{-1}}{\epsilon^{\prime}-(1+c)\epsilon} \ .
\end{align}
Optimized over $c$, and $(\eta,b)$,  this bound implies the following theorem which is the second main result of this section, where the optimum $c=\frac{(\epsilon^{\prime}-\epsilon)}{(\epsilon^{\prime}+\epsilon)}$.
\begin{theorem}\label{thm:CQ1_achie_2}
For a given a CQ channel $\cqN = \mathset{\rho_x}_{{x\in G}} $, and any $J$ as given in \eqref{eqn:J},
there exists a $(1,|J|, \epsilon^{\prime})$ group transmission system such that  
    \begin{align*}
      R \ge  \max_{\eta,b} \DHy^{\epsilon }(\pi^{AB}_{\eta,b}  \| \pi^{{AB}}_{\eta,b,J} )    - \log_2
  \frac{4\epsilon^{\prime}}{(\epsilon^{\prime}-\epsilon)^2},
    \end{align*} for any $\epsilon \in (0,\epsilon^{\prime})$, where the rate $R=\log_2 |J|$.
\end{theorem}

\subsection{Converse}\label{subsec:CQ1shot_conv}
Toward the converse, we have the following theorem whose proof is given in Appendix \ref{app:CQ_Error_C}.
Given a CQ channel 
$\mathcal{N}=\{\rho_x\}_{x \in G}$, and an input group $J$ as defined in \eqref{eqn:J}, 
and a $(1,|J|,\epsilon)$ group transmission system, 
define the input distribution $P_X$ by the uniform distribution on $\{\phi(J)+V\}$, where $\phi$ is the homomorphism and $V$ is the shift. 
Define the CQ state as 
\begin{align*}
  \rho^{{AB}}  \triangleq  \sum_{x} P_X(x)
  |x\ket\bra x|^A\otimes\rho_x^B.
\end{align*}
Define $\rho^{[A]B}$ and $\rho^{\bar{A}}$
 similar to those in Definition \ref{def:CQ_H_theta} for every $\hat{\theta} \neq \bs$ by noting that
$[X]$ is a function of $X$. 
\begin{theorem}\label{thm:CQ1_conv}
Given a CQ channel  $\mathcal{N}=\{\rho_x\}_{x \in G}$, and $J$ as given in \eqref{eqn:J}, 
   any $(1, \abs{J}, \epsilon)$ group transmission system satisfies
    % Assume that a group transmission system with parameters $(1, \abs{J}, \epsilon)$ exists  over a classical-quatum channel $\cqN = \mathset{\rho_x^B}_{{x\in \cX}}$, and that the group $J$ takes the form as in equation \eqref{eqn:J}. Then the rate of the code, $R = \log \abs{J}$, is bounded as:
    \[
    R \leq \min_{\hat{\theta}\neq \bs} \frac{1}{1- \omega_{\hat{\theta}}  } 
    I_{\textnormal{H}}^{\epsilon}(\overline{X}_{\eta,b}; [X_{\eta,b}],Y),
    % \DHy^\epsilon(\pi^{{AB}}  \| \pi^{{AB}}_{\hat{\theta}} ).
    \]
    where the rate $R=\log_2 |J|$.
\end{theorem}

\section{Group Coding in the Asymptotic Regime}
\label{sec:CCnshot}

The one-shot results can be used to derive a single-letter lower bound on the asymptotic group capacity of discrete memoryless channels using the AEP. This is accomplished in this section. A matching upper bound is also derived in the case of CQ channels.
For the classical channels, we use the upper bound available in \cite{sahebi2015abelian}.

\subsection{Classical Case}
% \blue{some duplicate def of Section \ref{subsec:H_coset}}
Consider a channel $(\mathcal{X}=G,\mathcal{Y},W_{Y|X})$. For every
$\eta$ and $b$, let $X_{\eta,b}$ be a random variable distributed
uniformly over $H_{\eta}+b$. Let
$[X_{\eta,b}]_{\hat{\theta}}=X_{\eta,b}+H_{\eta+\hat{\theta}}$ which
takes values from the cosets of $H_{\eta+\hat{\theta}}$ in $H_{\eta}+b$.
We leverage the one-shot results for classical channel group transmission and show the following capacity result. 
Let $\alpha$ be a probability distribution on the set of all pairs of $\eta$ and $b$,
where $\alpha_{\eta,b}$ denote the probability assigned to a particular
pair $(\eta,b)$. Recall from \eqref{eqn:J} that the input group $J$ can be specified by the vector $w$, which was fixed in the one-shot approach. In the asymptotic case, we optimize over all $J$.  

\begin{theorem}
\label{thm:CC_nshot_cap}
For a classical channel $(G,\mathcal{Y},W_{Y|X})$, the group capacity is given by
\begin{align}\label{eqn:Rate_Channel}
C= \sup_{\alpha,w} \min_{\hat{\theta} \neq \bs}
\frac{1}{1-\omega_{\hat{\theta}}} \sum_{\eta,b} \alpha_{\eta,b} I(X_{\eta,b};Y|[X_{\eta,b}]_{\hat{\theta}}).
\end{align}
\end{theorem}

\proof The proof is provided in Appendix~\ref{app:forCCnshot}.
\endproof

\begin{remark}
In the proof, we use the one-shot achievability bound given in Theorem \ref{thm:CC1_achieve} and apply on the $n$ uses of the given channel for asymptotically large $n$, and use the AEP to  obtain a lower bound on the group capacity of the channel.
We cannot directly use the one-shot converse here as the characterization given in Theorem \ref{thm:CC_nshot_cap} is single-letter in nature. This is an important issue that we want to emphasize. One-shot results are useful toward characterization of the asymptotic capacity only in terms of achievability. We appeal to the asymptotic converse coding theorem given in \cite{sahebi2015abelian}.
As stated in the introduction, the asymptotic performance limits of Abelian group codes have been characterized for classical discrete memoryless channels in \cite{como2009capacity,sahebi2015abelian} using varaints of mutual information under maximum likelihood or joint typical decoding procedures. 
Here we are rederiving the asymptotic result from the one-shot result.
\end{remark}

\subsection{Classical Quantum Case}

We provide a single-letter characterization of the group capacity for the CQ channel as follows. This is a new result. 
\begin{theorem}\label{thm:CQ_nshot_cap}  
For a CQ channel $\mathcal{N}=\{\rho_x\}_{x \in G}$, the group capacity is given by
\begin{equation*}
 C(\mathcal{N})= \sup_{\alpha,w} \min_{\hat{\theta} \neq \bs}
\frac{1}{1-\omega_{\hat{\theta}}} \sum_{\eta,b} \alpha_{\eta,b} I(X_{\eta,b};Y|[X_{\eta,b}]_{\hat{\theta}}).  
\end{equation*}
where 
\begin{align*}
   I(X;Y | [X]_{\hat{\theta}})
 \triangleq D(\rho^{A B} || \rho^{\bar{A}_{\hat{\theta}} } \rho^{[A]_{\hat{\theta}} B} ) \ .
\end{align*}
\end{theorem}
\proof The proof is provided in Appendix Section~\ref{app:forCQnshot}. The achievability follows by using Theorem \ref{thm:CQ1_achieve_1} and the AEP. 
A new asymptotic converse is developed to yield the single-letter characterization. 
\endproof

\section{Examples}
\label{sec:examples}

\begin{example}
We start with the classical binary symmetric channel with crossover probability $p$ and with input alphabet $G=\mathbb{Z}_2$. 
$J=\mathbb{Z}_2$. There are no non-trivial subgroups of $J$, implying that $\hat{\theta}=0$ and  $\eta=0$. 
Hence $w_{\hat{\theta}}=1$. The one-shot information rate is given by 
$I_{\textnormal{H}}^{\epsilon}(X;Y)$, which can be computed as follows:
\begin{align*}
I_{\textnormal{H}}^{\epsilon}(X;Y) &= \left\{
\begin{array}{cc}
0 & \mbox{if } \ \ 0 < \epsilon < p/2 \\
\log_2(4/3) & \mbox{ if }  \ \ p/2 \leq \epsilon <p \\
\log_2(2) & \mbox{ if  } \ \ p \leq \epsilon < (1+p)/2 \\
\log_2 4 & \mbox{ if  } \ \ (1+p)/2 \leq \epsilon <1
\end{array}
\right. .
\end{align*}
The asymptotic performance limit is the group capacity and is given by $I(X;Y)=1-h_b(p)$, where $h_b(\cdot)$ is the binary entropy function.
\end{example}

\begin{example}
    Consider a CQ channel given by $\mathcal{X}=\{0,1\}$, and 
    $\rho_{0}$ is given by the ensemble $(|0\ket,|1\ket)$, $(1/2,1/2)$, and
    $\rho_1$ is given by $(|+\ket,|-\ket)$, 
    $(1/4,3/4)$, where the states are the eigenvectors of Pauli $X$ operator. We have $G=J=\mathbb{Z}_2$. 
    As in the previous example, $\theta=0$ and $\eta=0$. The information rate is given by 
    $I_{\textnormal{H}}^{\epsilon}(X;Y)$, which can be computed as follows. The joint quantum state of the input and the output of the CQ channel is given by  $\rho^{AB}=\sum_{x \in \mathcal{X}} |x\ket \bra x|^A \otimes \rho^B_x$, and 
    the state $\rho^A=I/2$. 
The one-shot information rate can be evaluated numerically and is plotted in Figure \ref{fig:HT_mutual_info_binary}.
\begin{figure}[h!]
\begin{center}
  \includegraphics[width=3in]{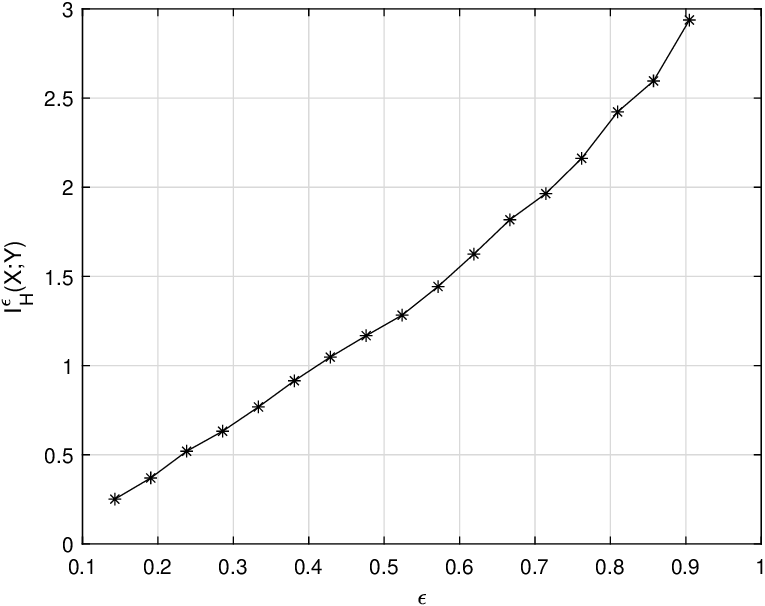}
  \end{center}
\caption{Numerical evaluation of the one-shot quantum information rate $I_{\textnormal{H}}^{\epsilon}(X;Y)$ for certain range of values of $\epsilon$.}
\label{fig:HT_mutual_info_binary}
\end{figure}
    The group capacity is  given by 
    \[
    I(X;Y)= H\left(0.5 (\rho_0+\rho_1)\right)-0.5 H(\rho_0)-0.5 H(\rho_1)=h_b(3/8)-1/2-1/2h_b(1/4). 
    \]    
\end{example}

\begin{example}
    Consider a classical octonary channel $Y=X \oplus_8 Z$, where $G=\mathbb{Z}_8$ and $P(Z=0)=1/2$ and $Z$ has the distribution: 
    $P_Z(i)=1/14$, for $i \neq 0$. Let the input group be $J=\mathbb{Z}_4$. We consider the case when $\eta=0$. There are two cases to consider: $\hat{\theta}=(1,0,3)$ and 
    $\hat{\theta}=(1,1,3)$. 

    Case 1: $\hat{\theta}_{2,2}=0$. $|H_{\hat{\theta}}|=|2\mathbb{Z}_8|=4$ and $|T_{\hat{\theta}}|=2$. We note that 
    $P_{XY}(x,y)=\frac{1}{8}P_Z(y-x)$ for all 
    $x,y \in G$. Moreover, we have for $i \in \{0,1\}$, $j, y \in G$, 
    \[
    P_{[X]}(i)P_{X|[X]}(j|i)P_{Y|[X]}(y|i) =\left\{ \begin{array}{cc} 
    \frac{1}{32}P_Z(y-i-
    2\mathbb{Z}_8) & \mbox{ if  }  (j-i)  \mbox{ is even } \\
    0 & \mbox{ otherwise}
    \end{array}.
    \right. 
    \]

Case 2: $\hat{\theta}_{2,2}=1$. $|H_{\hat{\theta}}|=|4\mathbb{Z}_8|=2$ and $|T_{\hat{\theta}}|=1$. We note that 
     for $i \in \{0,1,2,3\}$, $j, y \in G$, 
    \[
    P_{[X]}(i)P_{X|[X]}(j|i)P_{Y|[X]}(y|i) =\left\{ \begin{array}{cc} 
    \frac{1}{16}P_Z(y-i-
    4\mathbb{Z}_8) & \mbox{ if  }  4 \mbox{ divides } (j-i)  \\
    0 & \mbox{ otherwise}
    \end{array}
    \right. .
    \]
For several values of $\epsilon$, we have computed $I_{\textnormal{H}}^{\epsilon}(\overline{X};[X]Y)$
in Figure \ref{fig:octonary}.

\begin{figure} [h]
\begin{center}
  \includegraphics[width=3in]{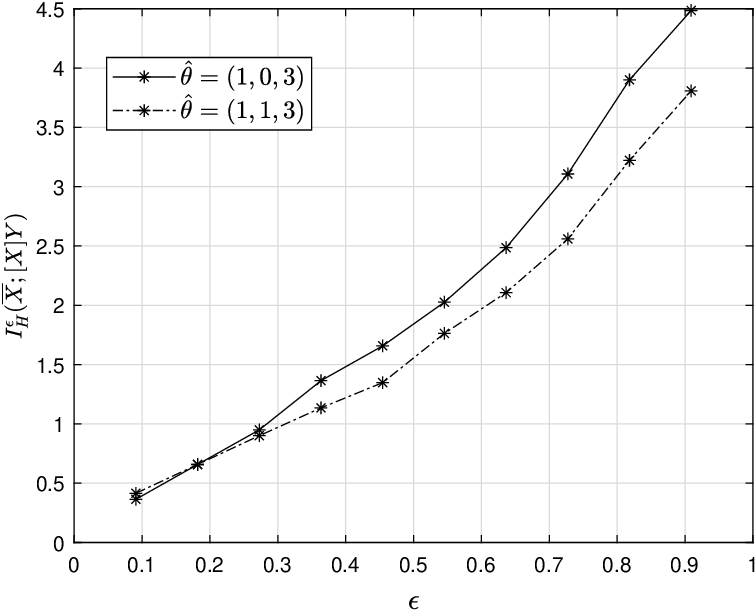}
  \end{center}
\caption{Numerical evaluation of the one-shot quantum information rate $I_{\textnormal{H}}^{\epsilon} (\overline{X};[X]Y)$ for certain range of values of $\epsilon$: $\hat{\theta}_{22}=0$ and $\hat{\theta}_{22}=1$.}
  \label{fig:octonary}
    \end{figure}    
\end{example}

\begin{example}
    \textbf{Coding over decimal numbers \{0,1,2\ldots 9\}:} Consider coding for a classical channel that takes a decimal number as input and produces a decimal number as output. Let us consider a symmetric model where $P_{Y|X}(y|x)=(1-p)$ if $x=y$ and $P_{Y|X}(y|x)=p/9$ if $x \neq y.$ The capacity of the channel is 
    given by $C(p)=\log(10)-h_b(p)-p\log(9)$. The only Abelian group on this alphabet is $\mathbb{Z}_{10}$, and  is isomorphic to 
    $\mathbb{Z}_2 \oplus \mathbb{Z}_5$. It has two non-trivial subgroups $\{0,5\}$ and $\{0,2,4,6,8\}$ corresponding 
to $\hat{\theta}=(0,1)$ and $\hat{\theta}=(1,0)$. One can evaluate a lower bound on the group capacity by choosing $\eta=0$
 to yield the following  function of $p$: 
 \[
 \sup_{0\leq w \leq 1} \min\{A,B,C\},
 \]
where 
\[
    A(p,w)=\frac{w+(1-w)\log(5)}{w} [h_b(8p/9)+24p/9+(1-8p/9)-h_b(p)-p \log(9)],
\]
\[
    B(p,w)=\frac{w+(1-w)\log(5)}{(1-w)\log(5)} [h_b(5p/9)+\log(5)-h_b(p)-p \log(9)].
\]
It turns out (which can be checked numerically) that this lower bound is indeed equal to the Shannon capacity. That is, $C = C(p)$ is smaller than both $A(p,w)$ and $B(p,w)$. 
This implies that the group capacity is equal to the Shannon capacity of the channel. Next we consider the largest rate achievable using linear codes over $\mathbb{Z}_7$ (the largest scalar finite field inside $\mathbb{Z}_{10}$) by using only the first $7$ symbols. The rate achievable using linear codes can be computed as 
\[
L(p) \triangleq [h_b(p/3)+(p/3)\log(3)+(1-p/3) \log(7)-h_b(p)-p \log(9)].
\]
The capacity  $C(p)$ and the rate $L(p)$ are plotted in Figure \ref{fig:z10}. For the symmetric channel, linear codes cannot achieve the capacity. 
\begin{figure} [h]
\begin{center}
  \includegraphics[width=3in]{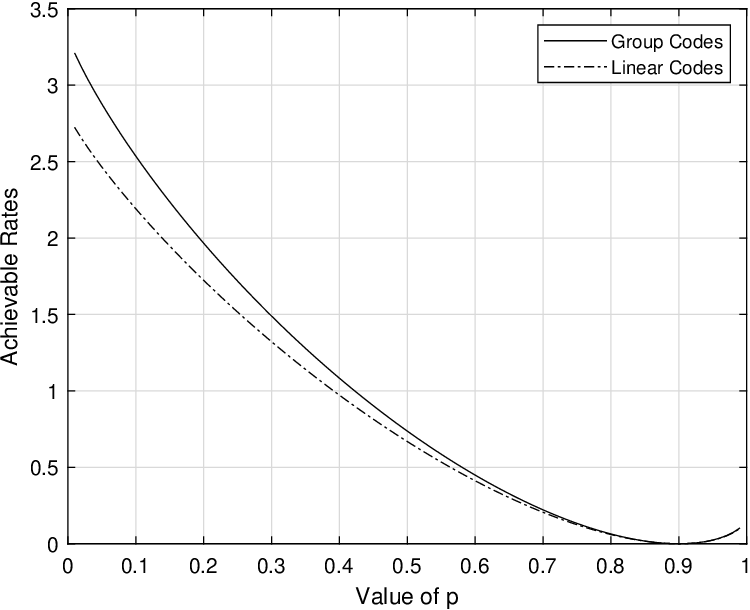}
  \end{center}
\caption{Codes over the decimal numbers for a symmetric channel: Group codes achieve the capacity while linear codes over $\mathbb{Z}_7$ do not.  }
  \label{fig:z10}
    \end{figure}

\end{example}

\begin{example}
    Consider a CQ channel $\mathcal{N}=\{\rho_x\}_{x \in \mathcal{X}}$ with quaternary input 
    $\mathcal{X}=\{0,1,2,3\}$ with $G=\mathbb{Z}_4$. We compute the group capacity of this channel in the asymptotic regime. $G^n=\mathbb{Z}_{4}^n$ and 
    $J=\mathbb{Z}^{k_1}_2 \oplus \mathbb{Z}^{k_2}_4$. We have $w_1+2w_2=1$,
    $w_1=\frac{k_1}{k_1+k_2}$, and 
    $w_2=\frac{k_2}{k_1+k_2}$. 
    There are $5$ non-trivial vectors $\eta$: $(0,0),(0,1),(1,0),(1,1)$, and $(0,2)$. which yield 
    $H_{\eta}$ as follows: $\mathbb{Z}_4$
 in the first $3$ cases and $2\mathbb{Z}_4$ in the next $2$ cases. 
 There are two non-trivial cases of vector $b$ (which is a scalar for the present example): $b=0$ and  $b=1$. Recall that 
 $I(X;Y | [X]_{\hat{\theta}})
 = D(\rho^{A B} || \rho^{\bar{A}_{\hat{\theta}} } \rho^{[A]_{\hat{\theta}} B} ).$
 Define 
 $I_4:=I(X;Y)$, $I_2=I(X;Y|[X]=0)$ and 
 $I_2'=I(X;Y|[X]=1)$. 
 The capacity from Theorem \ref{thm:CQ_nshot_cap} can be expressed as
 \[
C(\mathcal{N})= \sup_{\alpha_0,\alpha_1,\alpha_2,w_2} \min\{A_1,A_2,A_3\},
 \]
where $0\leq \alpha_0,\alpha_1,\alpha_2,w_2 \leq 1$ and $\alpha_0+\alpha_1+\alpha_2=1$, and 
\[
A_1=\alpha_0 I_4+\alpha_1 I_2+\alpha_2 I_2',  \ \
A_2=(1+w_2)[\frac{\alpha_0}{2} (I_2+I_2')+\alpha_1 I_2+\alpha_2 I'_2], \ \ 
A_3=\frac{(1+w_2)}{w_2} \frac{\alpha_0}{2} (I_2+I_2'). 
\]
Since the expression is symmetric in $I_2$ and $I_2'$, without loss of generality, let us assume that $I'_2 \leq I_2$. There are three cases to consider. Case (a) $I_4 \leq I_2$: Then $\min\{A_1,A_2,A_3\} \leq A_1\leq I_2$. The upper bound can be reached by $\alpha_1=1$ and any $w_2$. Case (b): $I_2 \leq I_4 \leq (I_2+I_2')$: Then 
$\min\{A_1,A_2,A_3\} \leq A_1\leq I_4$. The upper bound can be reached by $\alpha_0=1$ and $w_2=1$.  Case (c): $I_2+I'_2\leq I_4$: 
Then $\min\{A_1,A_2,A_3\} \leq A_2 \leq I_2+I_2'$. The upper bound can be reached by 
$\alpha_0=1$ and $w_2=1$.
Hence the expression for the group capacity can be simplified as 
\[
C(\mathcal{N})=\max\{ \min\{I_4,I_2+I'_2\},I_2,I_2'\}.
\]
This corresponds to three possibilities: (a) use the entire channel input alphabet $\mathbb{Z}_4$ to get the first expression inside the maximization or (b) use only a subset $2\mathbb{Z}_4$,
or $2\mathbb{Z}_4+1$ to get the other two expressions, respectively. 
Let us take a particular example of CQ channel.    Let $\rho_0$ be given by the ensemble $(|0\ket,|1\ket)$, $(1/2,1/2)$, $\rho_1$ be given by 
    $(|+\ket,|-\ket)$, $(1/4,3/4)$,
    $\rho_2$ by  $(|+\ket,|-\ket)$, $(3/4,1/4)$,
    and $\rho_3$ by $(|i\ket,|-i\ket)$,
    $(1/4,3/4)$, where the last set of states are the eigenvectors of Pauli-Y operator.
The quantum mutual information quantities can be computed as $I_4=0.1302$, $I_2=0.0488$ and 
$I'_2=0.0966$. The capacity is $C(\mathcal{N})=0.1302$.    
\end{example}

\begin{example}
    \textbf{Prime-based linear codes:} Let us consider a channel 
    (either classical or classical-quantum) with a prime input alphabet $\mathcal{X}=\{0,1,\ldots,p-1\}$ for a prime $p$.  Consider $n$ uses of the channel
    for some finite $n$. For this case, we let $G=\mathbb{Z}_p^n$ and $J=\mathbb{Z}_p^k$ for some $k$.
$\mathcal{P}(G)=\{p\}$, $\mathcal{R}_2(G)=\{1\}$, and 
$S(G)=\{(p,1)\}.$ There is only one choice: $\hat{\theta}_{p,1}=0$, and $w_{\hat{\theta}}=0$. 
Moreover, we have $\eta=0$.  This corresponds to the standard binary $(n,k)$ linear code. 
The one-shot information rate is 
given by $I_{\textnormal{H}}^{\epsilon}(X^n;Y^n)$. 
One can check that this is also the one-shot information rate for random unstructured codes with 
uniform distribution on the input alphabet $\mathcal{X}^n$.  
This demonstrates that random linear code ensemble  achieves the same performance as the random unstructured code ensemble
with uniform distribution for both classical and classical-quantum channels for any $n$.
\end{example}

\bibliographystyle{IEEEtran}
\bibliography{IEEEabrv}
%%
%% where we here have assume the existence of the files
%% definitions.bib and bibliofile.bib.
%% BibTeX documentation can be obtained at:
%% http://www.ctan.org/tex-archive/biblio/bibtex/contrib/doc/
%%%%%%
%\newpage
 % \,
 % \newpage

\appendix
\subsection{More on the $H_{\hat{\theta}}$ coset}\label{appendix:H_coset}
We show the derivation for \eqref{eq:T_theta_size} as follows: 
\begin{align}
\log \abs{T_{\hat{\theta}}}
&= \sum_{(p,s)\in\mathcal{S}(G)} {(s- \hat{\theta}_{p,s}) k_{p,s}} \log p \nonumber \\
&= k \cdot \sparenv{\frac{ \sum_{(p,s)\in\mathcal{S}(G)} {(s- \hat{\theta}_{p,s}) \omega_{p,s}} \log p }{ \sum_{(p,s)\in\mathcal{S}(G)} {s \omega_{p,s}} \log p}}   \sum_{(p,s)\in\mathcal{S}(G)} {s \omega_{p,s}} \log p \nonumber \\
&= (1- \omega_{\hat{\theta}} )k \sum_{(p,s)\in\mathcal{S}(G)} {s \omega_{p,s}} \log p = (1- \omega_{\hat{\theta}} ) nR. \label{eq:T_theta_size_2}
\end{align}

We recall the following result from \cite{sahebi2015abelian}. 
\begin{lemma}\label{lem:Pr_Ttheta}
For $a,\tilde{a}\in J$, $x,\tilde{x}\in G^n$ and for $(p,s)\in\mathcal{Q}(J) = \mathcal{S}(G)$, let $\hat{\theta}_{p,s}\in\{0,1,\cdots,s\}$ be such that
% \begin{align*}
$\tilde{a}_{p,s} - a_{p,s}\in p^{\hat{\theta}_{p,s}}\mathds{Z}_{p^s}^{k_{p,s}}
\backslash
p^{\hat{\theta}_{p,s}+1}\mathds{Z}_{p^s}^{k_{p,s}}, 
$ i.e., 
$\tilde{a} \in  T_{\hat{\theta}}(a)$.
% \end{align*}
Consider a random homomorphism $\phi$ and a dither $V$ with distribution specified as in Definition~\ref{def:groupCode}. Then,
\begin{align*}
\Pr\left({ \phi(a)+V=x,  \phi(\tilde{a})+V=\tilde{x} }\right)
= \left\{\begin{array}{ll}
\frac{1}{|G|^n}\frac{1}{|H_{\hat{\theta}}|^n} &\mbox{if $\tilde{x}-x\in H_{\hat{\theta}}^n$}\\
0 &\mbox{otherwise}
\end{array}\right.
\end{align*}
\end{lemma}

\subsection{Proof of Achievability for Classical Channel}
\label{app:CC_Error}
Denote by $P_{err}(u)$ the probability of the event $E_2(u)\cap (E_1(u))^c$, averaged over the randomness of $\phi, V$.  
We can provide an upper bound on this probability as 
\begin{align}
% \label{eq:Pe_u_exmaple1}
P_{err}(u) &\triangleq\, 
\mE\sparenv{ \Pr(E_2(u)\cap (E_1(u))^c)} \nonumber \\
&= \sum_{(x,y)\in A_{\epsilon}} W(y|x)
\Pr\left(
\substack{
\phi(u)+V=x, \exists \tilde{u}\in J: \\ \tilde{u}\ne u, (\phi(\tilde{u})+V, y) \in A_{\epsilon} }
\right)\nonumber \\
&\leq 
\sum_{(x,y)\in A_{\epsilon}} 
\sum_{\substack{
\tilde{u}\in J\\ \tilde{u}\neq u}}
\sum_{\substack{\tilde{x}\in G\\(\tilde{x},y)\in A_{\epsilon}}}
W(y|x)
\Pr\left(
\substack{
\phi(u)+V=x, \\ \phi(\tilde{u})+V=\tilde{x}
}
\right)\nonumber \\
&= \sum_{{\hat{\theta}} \neq \bs }
\sum_{(x,y)\in A_{\epsilon}} 
\sum_{
\tilde{u}\in T_{{\hat{\theta}}}(u)}
\sum_{\substack{\tilde{x}\in G\\(\tilde{x},y)\in A_{\epsilon}}}
W(y|x)
\Pr\left( \substack{
\phi(u)+V=x, \\ \phi(\tilde{u})+V=\tilde{x}
} \right) \nonumber \\
&= \sum_{{\hat{\theta}} \neq \bs } 
P_{err}(u, \hat{\theta}), \label{eq:Perr_utheta_CC-1}
\end{align}
where 
$P_{err}(u, \hat{\theta}) \triangleq \,
\sum_{
\tilde{u}\in T_{{\hat{\theta}}}(u)}
P_{err}(u, \tilde{u})$ 
and for 
$\tilde{u} \in T_{{\hat{\theta}}}(u)$, 
% \begin{align}
\begin{equation} \label{eq:Perr_uutilde_CC1}
P_{err}(u, \tilde{u}) \triangleq \sum_{(x,y)\in A_{\epsilon}} 
\sum_{\substack{\tilde{x}\in G\\(\tilde{x},y)\in A_{\epsilon}}}
 W(y|x)  
\Pr\left( \substack{\phi(u)+V=x, \\ \phi(\tilde{u})+V=\tilde{x}}
\right),   
\end{equation}
% \end{align}
and $\bs$ denote the vector whose components satisfy $\bs_{(p,s)} = s$ for all $(p,s)\in \mathcal{S}(G)$.
The term $\Pr(\phi(u)+V=x,$ $\phi(\tilde{u})+V=\tilde{x})$ in \eqref{eq:Perr_uutilde_CC1} can be found using Lemma~\ref{lem:Pr_Ttheta}. 
Let $x \in \sparenv{x_r}$ be a shorthand for $[x] = \sparenv{x_r}$, or equivalently, $x \in \sparenv{x_r}+ H$.
Hence  
\begin{align}
P_{err}(u, \tilde{u}) 
&=
\sum_{(x,y)\in A_{\epsilon}} 
\sum_{\tilde{x} \in x+H_{\hat{\theta}} }
W(y|x) \ind_{A_{\epsilon}}(\tilde{x},y)
\frac{1}{\abs{G}}\frac{1}{\abs{H_{\hat{\theta}}}}  \nonumber \\
&= 
\sum_{x \in \cX} 
\sum_{\tilde{x} \in x+H_{\hat{\theta}} }
\sum_{\substack{ y: ({x}, y)\in A_{\epsilon}\\(\tilde{x}, y)\in A_{\epsilon} }}
W(y|x) 
\frac{1}{\abs{G}}\frac{1}{\abs{H_{\hat{\theta}}}} \nonumber \\
&\leq 
\sum_{x \in \cX} 
\sum_{\tilde{x} \in x+H_{\hat{\theta}} }
\sum_{y: (\tilde{x}, y)\in A_{\epsilon} }
W(y|x) 
\frac{1}{\abs{G}}\frac{1}{\abs{H_{\hat{\theta}}}} \nonumber \\
&= 
\sum_{\sparenv{x_r}} \frac{\abs{H_{\hat{\theta}}}}{\abs{G}}
\sum_{x \in \sparenv{x_r}} 
\sum_{\tilde{x} \in \sparenv{x_r}}
\sum_{y: (\tilde{x}, y)\in A_{\epsilon} }
W(y|x) 
\frac{1}{\abs{H_{\hat{\theta}}}^2} \nonumber \\
&=\sum_{\sparenv{x_r}} P(\sparenv{x_r}) 
\sum_{\tilde{x} \in \sparenv{x_r}}
\sum_{y: (\tilde{x}, y)\in A_{\epsilon} } P(\tilde{x} \ \vert \sparenv{x_r}) 
  P(y\ \vert  \sparenv{x_r})\, .
\label{eq:Perr_uutilde_CC1_expanded}
\end{align}
The probability of the event $E_1(u)$ can be bounded as: \begin{align*}
         \Pr(E_1(u)) 
         &= \Pr\parenv{ ( \phi(u) +V , Y)\notin A_{\epsilon} }
         = \Pr((X, Y)\notin A_{\epsilon}) \\
        &= \Pr\parenv{(X, Y)\in \cup_{\hat{\theta}} ({A^{\ast}_{\epsilon_{\hat{\theta}}}})^C} \leq \sum_{\hat{\theta}}\Pr\parenv{(X, Y)\in ({A^{\ast}_{\epsilon_{\hat{\theta}}}})^C} \\
        &\leq \sum_{\hat{\theta}}\epsilon_{\hat{\theta}} = \epsilon.
\end{align*}
Since $A_{\epsilon} \subset A^{\ast}_{\epsilon_{\hat{\theta}}}$, we may exploit \eqref{eq:Perr_uutilde_CC1_expanded} and show that the term $P_{err}(u, \tilde{u})$  is then bounded from above as follows: 
\begin{align*}
P_{err}(u, \tilde{u}) \leq &\sum_{\sparenv{x_r}} P(\sparenv{x_r}) 
\sum_{\tilde{x} \in \sparenv{x_r}}
\sum_{y: (\tilde{x}, y)\in A^{\ast}_{\epsilon_{\hat{\theta}}} } 
P(\tilde{x} \vert \sparenv{x_r}) P(y\vert  \sparenv{x_r}) =\exp_2 \Large\{ -I_{\textnormal{H}}^{\epsilon_{\hat{\theta}},\hat{\theta}}(\overline{X}; [X]Y) \Large\}, 
\end{align*}
which leads to the following bound, 
% implies the following bound on $P_{err}(u)$:
\begin{align}\label{eq:perr_CC1_achieve}
% $    
P_{err}(u) \leq \sum_{{\hat{\theta}} \neq \bs } \abs{T_{{\hat{\theta}}}(u)} 
\exp_2 \Large\{ -I_{\textnormal{H}}^{\epsilon_{\hat{\theta}},\hat{\theta}}(\overline{X}; [X]Y) \Large\}.
% $
\end{align} 
Therefore, we have 
 % \begin{align}
\begin{align*}
   \Pr(E(u)) \leq \epsilon + \sum_{{\hat{\theta}} \neq \bs }  \abs{T_{{\hat{\theta}}}(u)} \exp_2  \Large\{-I_{\textnormal{H}}^{\epsilon_{\hat{\theta}},\hat{\theta}}(\overline{X}; [X]Y) \Large\}.     
\end{align*}
 % \end{align}
The average probability of error of the group transmission scheme can be bounded from above by 
\begin{align*}
\Pr(\textnormal{error}) = \sum_{u\in J}\frac{1}{\abs{J}}  \Pr(E(u)) 
\leq  \epsilon + 
\sum_{{\hat{\theta}} \neq \bs }  \abs{T_{{\hat{\theta}}}(u)} \exp_2 
\Large\{-I_{\textnormal{H}}^{\epsilon_{\hat{\theta}},\hat{\theta}}(\overline{X}; [X]Y) \Large\}\, .
% \bracenv{ -D_H^{\epsilon_{\hat{\theta}}}(P_{XY} \| P_{[X]_{\hat{\theta}}} P_{X\mid [X]_{\hat{\theta}}} P_{Y\mid [X]_{\hat{\theta}}} ) }.  
\end{align*}

\subsection{Proof of Proposition \ref{prop:proposition}}
\label{app:prooof_proposition}
Recall that the set $A^{\ast}_{\epsilon_{\hat{\theta}}}$ in the proof of \Tref{thm:CC1_achieve} is the minimizer in the definition of the right-hand side of \eqref{eq:DHdef_classical} for 
$I_{\textnormal{H}}^{\epsilon_{\hat{\theta}}}(\overline{X}; [X]Y),$
% $D_H^{\epsilon_{\hat{\theta}}, \hat{\theta}} (P_{XY} \| P_{[X]} P_{X\mid [X]} P_{Y\mid [X]} )$
 i.e., $P_{XY}(A^{\ast}_{\epsilon_{\hat{\theta}}}) \geq 1- \epsilon_{\hat{\theta}}$ and
\begin{align*}
    I_{\textnormal{H}}^{\epsilon_{\hat{\theta}}}(\overline{X}; [X]Y)
    &= -\log_2  
\Big[   \sum_{\sparenv{x_r}} P(\sparenv{x_r}) 
    \sum_{x\in \sparenv{x_r}} P(x \,\vert \sparenv{x_r})
    \sum_{y: (x,y)\in A^{\ast}_{\epsilon_{\hat{\theta}}}}
    P(y\,\vert  \sparenv{x_r})\Big]\, 
    = -\log_2  
   P_{XY\mid \hat{\theta}}( A^{\ast}_{\epsilon_{\hat{\theta}}} ) \, .
\end{align*}
Denote by  $A^{\textnormal{T1}}_{\epsilon} $ the intersection  $A^{\textnormal{T1}}_{\epsilon} = 
\cap_{\hat{\theta} \neq \bs}  A^{\ast}_{\epsilon_{\hat{\theta}}}$. 
The probability of error for message $u$, denoted by $ \Pr^{\textnormal{T1}}(E(u))$, is bounded as $ \Pr^{\textnormal{T1}}(E(u)) \leq \epsilon + P_{err}(u), $
%  \begin{align}
%  \Pr(E(u)) \leq \epsilon + P_{err}(u), 
% \end{align} 
where $P_{err}(u)$ is  bounded in \eqref{eq:perr_CC1_achieve} as 
\begin{align*}
P_{err}(u)\leq 
\sum_{{\hat{\theta}} \neq \bs }  \abs{T_{{\hat{\theta}}}(u)} P_{XY\mid \hat{\theta}}( A^{\ast}_{\epsilon_{\hat{\theta}}} )
=\sum_{{\hat{\theta}} \neq \bs }  \abs{T_{{\hat{\theta}}}(u)} \exp_2 \bracenv{ -
I_{\textnormal{H}}^{\epsilon_{\hat{\theta}}} (\overline{X}; [X]Y)} 
 \,.
\end{align*}

In the proof of \Tref{thm:CC1_achieve_2nd}, the probability of error for message $u$, denoted here by $\Pr^{\textnormal{T2}}(E(u))$, is bounded as $ \Pr^\textnormal{T2}(E(u)) \leq \epsilon + P^{\textnormal{T2}}_{err}(u), $ where we have 
\begin{align*}
     P^{\textnormal{T2}}_{err}(u) \leq  
     \sum_{{\hat{\theta}} \neq \bs }  \abs{T_{{\hat{\theta}}}(u)} P_{XY\mid \hat{\theta}}( A^{\textnormal{T2}}_{\epsilon} )
     =
     ( \abs{J} -1)  P_{XY\mid J } (A^{\textnormal{T2}}_{\epsilon})\, . 
\end{align*}
Here the set $A^{\textnormal{T2}}_{\epsilon} $ is a subset of $\cX\times \cY$ that achieves  $D_{\textnormal{H}}^{\epsilon}(P_{XY} \| P_{XY\mid J })$. That is, 
$P_{XY\mid J } (A^{\textnormal{T2}}_{\epsilon} ) \leq  P_{XY\mid J }(A)$ for any $A$ such that 
$P_{XY}(A) \geq 1-\epsilon $. 

Note that $P_{XY}(A^{\textnormal{T1}}_{\epsilon}) \geq 1-\epsilon $. We thus have  $P_{XY\mid J } (A^{\textnormal{T2}}_{\epsilon} ) \leq  P_{XY\mid J }(A^{\textnormal{T1}}_{\epsilon})$ and 
\begin{align*}
     \sum_{{\hat{\theta}} \neq \bs }  \abs{T_{{\hat{\theta}}}(u)} P_{XY\mid \hat{\theta}}( A^{\textnormal{T2}}_{\epsilon} ) 
     \leq 
     \sum_{{\hat{\theta}} \neq \bs }  \abs{T_{{\hat{\theta}}}(u)} P_{XY\mid \hat{\theta}}( A^{\textnormal{T1}}_{\epsilon} ) 
     \leq 
     \sum_{{\hat{\theta}} \neq \bs }  \abs{T_{{\hat{\theta}}}(u)} P_{XY\mid \hat{\theta}}( A^{\ast}_{\epsilon_{\hat{\theta}}} )\, . 
\end{align*}
Therefore the upper bound on $ \Pr^{\textnormal{T2}}(E(u))$ is smaller or equal to that on $ \Pr^{\textnormal{T1}}(E(u))$.
This completes the proof.

\subsection{Proof of Converse for Classical Channel}
\label{app:CC_Error_C}
Based on Lemma~\ref{lem:homo_JtoG}, for each group code $\mathds{C}\leq G$, there exists a group $J$ and a homomorphism such that $\mathds{C}$ is the image of the homomorphism. 
Assume now that a group transmission system with parameters $(1, \abs{J}, \epsilon)$ exists over a channel $(\mathcal{X}=G,\mathcal{Y},W_{Y|X})$, and that the group $J$ takes the form in equation \eqref{eqn:J}. 
{Assume that the homomorphism $\phi$ for the group code $\mathds{C}$ is a one-to-one mapping. }
We have: 
\[\mathds{C} =  \big\{ \bigoplus_{(p,r,m)\in \mathcal{G}(G)} \sum_{s=1}^{r_p} 
{u}_{p,s} g_{(p,s)\rightarrow(r,m)} +V: u\in J
\big\}.
\]

Let $\hat{\theta}$ be a vector indexed by $(p,s)\in \mathcal{S}(G)$ with $0\leq \hat{\theta}_{p,s}\leq s$.
For a message $u\in J$,
% =\bigoplus_{p\in\mathcal{P}(G)} \bigoplus_{s=1}^{r_p} \mathds{Z}_{p^s}^{k_{p,s}} $
construct a one-to-one correspondence between $u_{p,s}\in \mathds{Z}_{p^s}^{k_{p,s}}$ and the tuple $(\tilde{u}_{p,s}, \hat{u}_{p,s})$ where $\tilde{u}_{p,s}\in p^{\hat{\theta}_{p,s}} \mathds{Z}_{p^s}^{k_{p,s}}$ and $ \hat{u}_{p,s} \in \mathds{Z}_{p^{\hat{\theta}_{p,s}}}^{k_{p,s}}$.
Let $U$ denote the random message of the group transmission system of the code. Let $\hat{U}$ denote the part of the random message such that $\hat{U}_{p,s} \in \mathds{Z}_{p^{\hat{\theta}_{p,s}}}^{k_{p,s}}$, for all $(p,s)\in \mathcal{S}(G)$. 
Consider the subcode of $\mathds{C}$: 
\begin{align*}
    \mathds{C}_1(\hat{\theta}, \hat{u}) = \big\{ \bigoplus_{(p,r,m)\in \mathcal{G}(G)} \sum_{s=1}^{r_p} 
(\tilde{u}_{p,s} +\hat{u}_{p,s}) g_{(p,s)\rightarrow(r,m)} +V:
\tilde{u}_{p,s}\in p^{\hat{\theta}_{p,s}} \mathds{Z}_{p^s}^{k_{p,s}}, \forall (p,s)\in \mathcal{S}(G)
\big\}.
\end{align*}
Let $x = \phi(u) +V$ be the channel input and $H_{\hat{\theta}}$ be given as in \eqref{eqn:H_theta}. Then $\mathds{C}_1(\hat{\theta}, \hat{u}) = x+H_{\hat{\theta}}$. That is, there is a one-to-one correspondence between $\hat{U}$  and $[X]_{\hat{\theta}}$. 
Also, $\abs{T_{\hat{\theta}}(u)} = \abs{\mathds{C}_1(\hat{\theta}, \hat{u}) } = \abs{H_{\hat{\theta}}}$ for all $u\in J$.
Let $x \in \sparenv{x_r}$ be a shorthand for $[x] = \sparenv{x_r}$, or equivalently, $x \in \sparenv{x_r}+ H$, where we omit the $\hat{\theta}$ subscript when it is clear from the context.

Define a one-to-one correspondence between $x$ and the tuple $(\tilde{x}_{\hat{\theta}}, [x]_{\hat{\theta}})$ where $\tilde{x}_{\hat{\theta}} = \phi(\tilde{u})$.
Consider a genie-aided receiver which gets access to  $\hat{U}$ and performs maximum likelihood decoding. Equivalently, this receiver has access to the coset information $[X]_{\hat{\theta}}$ of $X$ and can be written as $\sD^{ga}:([x]_{\hat{\theta}}, y) \mapsto x' \in \cX$. Clearly the average probability of error for this decoder must be not greater than $\epsilon$. Let $X' \in \cX$ be the output of $\sD^{ga}$. 
For every $\hat{\theta}$ with $0\leq \hat{\theta}_{p,s}\leq s$, $\hat{\theta}\neq \mathbf{s}$, the average probability of error for this decoder is 
\begin{align*}
&\sum_{\hat{u}} \sum_{x,x'} \Pr(\hat{u}) P_{XX'\mid \hat{U}}(x, x'\mid \hat{u})\one_{ \mathset{x' \neq x}}  = \sum_{x,x'} P_{XX'}(x,x')\one_{ \mathset{x' \neq x}}   \leq \epsilon,
\end{align*} 
where 
\[
P_{XX'\mid \hat{U}}(x, x'\vert \hat{u})\triangleq P(x\vert [x]) \sum_{y: \sD^{ga}([x], y) = x' } W(y\vert x)\, .
\]

Consider a strategy to distinguish between   $P_{XX'}$ and $P_{\hat{U}} (P_{X\mid \hat{U}} \otimes P_{X' \mid \hat{U}})$ as follows. The strategy guesses $P_{XX'}$ if it sees $X = X'$, and guesses  $P_{\hat{U}} (P_{X\mid \hat{U}} \otimes P_{X' \mid \hat{U}})$ otherwise.
When $P_{XX'}$ is the true underlying distribution, the type-I error probability is exactly the probability that $X\neq X'$ computed from $P_{XX'}$, namely, the average probability of a decoding error, and is thus not larger than $\epsilon$. 
When $P_{\hat{U}} (P_{X\mid \hat{U}} \otimes P_{X' \mid \hat{U}})$ is the true underlying distribution, the probability of type-II error (misdetection) is 
\begin{align}
\sum_{\hat{u}} P_{\hat{U}}(\hat{u}) 
\sum_{x,x'}  P_{X\mid \hat{U}} (x\mid\hat{u})  P_{X' \mid \hat{U}}(x'\mid\hat{u})\one_{ \mathset{x' = x}} 
&=\sum_{[x_r]} P([x_r]) 
\sum_{x,x'}  P(x\mid [x_r])  P(x'\mid [x_r]) \one_{ \mathset{x' = x}} \label{eq:prob_misd_2} \\
& =\sum_{[x_r]} \frac{\abs{H}}{\abs{G}} 
\sum_{x}  P (x\mid [x_r])  P(x \mid [x_r]) \nonumber \\
&= \sum_{[x_r]} \frac{\abs{H}}{\abs{G}} 
\sum_{x\in [x_r]}  \frac{1}{\abs{H}}  P(x \mid [x_r]) =  \frac{1}{\abs{H}},\nonumber 
% \label{eq:prob_misd_4}
\end{align} where \eqref{eq:prob_misd_2} follows from the one-to-one mapping between $\hat{U}$  and $[X]_{\hat{\theta}}$.
Thus, 
\begin{align}\label{eq:subgroup_relativeEntropy_CC1}
% &\Large\{-I_{H}^{\epsilon,\hat{\theta}}(\overline{X}; [X]Y) \Large\}\\
D_{\textnormal{H}}^{\epsilon,\hat{\theta}}(P_{XY} \| P_{[X]} P_{X\mid [X]} P_{Y\mid [X]} ) 
\geq
&D_{\textnormal{H}}^{\epsilon, \hat{\theta}}(P_{XX'} \| P_{[X]} P_{X\mid [X]} P_{X'\mid [X]} ) \nonumber \\
= &D_{\textnormal{H}}^{\epsilon}(P_{XX'} \| P_{\hat{U}} (P_{X\mid \hat{U}} \otimes P_{X' \mid \hat{U}}) ) \nonumber \\ 
\geq
&-\log_2 \frac{1}{\abs{H}} = \log_2 \abs{H} \nonumber = \log_2 \abs{T_{\hat{\theta}}(u)} \\
= &(1- \omega_{\hat{\theta}} )k \sum_{(p,s)\in\mathcal{S}(G)} {s \omega_{p,s}} \log p,
% &= \log_2 \prod_{(p,r,m)\in\mathcal{G}(G)} p^{r -\pmb{\theta}(\hat{\theta})_{p,r}} \nonumber  =\sum_{(p,r,m)\in\mathcal{G}(G)} \sparenv{r -\pmb{\theta}(\hat{\theta})_{p,r} } \log_2 p \\
% = \sum_{(p,r)\in \mathcal{Q}(G)} m\sparenv{r -\pmb{\theta}(\hat{\theta})_{p,r} } \log_2 p, 
\end{align}
where the first inequality follows from the data processing inequality \cite{wang2012one}.
% and the last equality is shown in equation \eqref{eq:T_theta_size} in Section~\ref{subsec:H_coset}.  
Equivalently, 
% we may rewrite \eqref{eq:subgroup_relativeEntropy_CC1} compactly as 
\begin{equation}
    I_{\textnormal{H}}^{\epsilon}(\overline{X}; [X]Y) 
% D_H^{\epsilon,\hat{\theta}}(P_{XY} \| P_{[X]} P_{X\mid [X]} P_{Y\mid [X]} )  
\geq (1- \omega_{\hat{\theta}} ) R\, , \label{eq:CC1converse_pf_zeroEta}
\end{equation}
which yields \Tref{thm:CC1_conv}.

\subsection{Proof of Achievability for CQ Channel }
\label{app:CQ_Error}
We work on the two terms in the right hand side of (\ref{eq:CQ_achieve_simplify}). The first term can be simplified as follows:
\begin{align*}
    \EE_{C} 
\trace &\sparenv{ \parenv{
    \one^{B'} - (\Pi')^{B'}_{([x], [\Bell])( [u]), (\overline{x},\bar{\Bell}) (u)} }({\rho'})^{B'}_{([x], [\Bell])( [u]), (\overline{x},\bar{\Bell}) (u), \delta }
 }\\
     = & \frac{1}{|\mathcal{L}|^{\zeta(G)}} \sum_{x,\Bell}
     P(x) 
        \trace   [ (\rho')^{B'}_{x,\Bell}-(\Pi')^{B'}_{x,\Bell}
   (\rho')^{B'}_{x,\Bell}]  \\
   \stackrel{(a)}{\leq} & \frac{4}{|\mathcal{L}|^{\zeta(G)}} \sum_{x,\Bell}
     P(x) 
   \trace   [(I- (\Pi)^{B'}_{\hat{B}})
   (\rho')^{B'}_{x,\Bell}+(\Pi')^{B'}_{W'_{x,\Bell}}
   (\rho')^{B'}_{x,\Bell}]  \\
   \stackrel{(b)}{\leq} & 16 \zeta(G)\delta^2+ \frac{4}{|\mathcal{L}|^{\zeta(G)}} 
\sum_{x,\Bell}
     P(x) 
   \trace [(\Pi')^{B'}_{W'_{x,\Bell}} (\rho^B_{x} \otimes |0 \ket \bra 0 | )] \\
\stackrel{(c)}{\leq} & 16 \zeta(G)\delta^2+ \frac{4(1+\zeta(G)\delta^2)}{|\mathcal{L}|^{\zeta(G)} \delta^2} 
\sum_{x,\Bell}
     P(x) 
       \sum_{\hat{\theta} \in \Theta}
  \left[1-   \trace [(\Pi)^{\hat{B}}_{[X];x} (\rho^B_{x} \otimes |0 \ket  \bra 0|)] \right]  \\
  = & 16 \zeta(G)\delta^2+ \frac{4(1+\zeta(G)\delta^2)}{|\mathcal{L}|^{\zeta(G)} \delta^2} 
\sum_{x,\Bell}
     P(x)  \sum_{\hat{\theta}\in \Theta} [1-   \trace [(\Pi'')^{B}_{[X];x} \rho^B_{x}]  \\ 
  \stackrel{(d)}{\leq} & 16\zeta(G)\delta^2 +\frac{4(1+\zeta(G)\delta^2) |\Theta| \epsilon}{\delta^2},
\end{align*}
where we provide the following arguments.
(a) follows from Fact 3 of \cite{sen2021unions} and  (b)  from (\ref{eq:closeness_rho_new}). (c) follows from 
Proposition 2 \cite{sen2021unions} by using
$l=|\Theta|$, and 
 $\alpha=\frac{\delta^2}{1+\zeta(G)\delta^2}$, and 
 (d) follows from 
 (\ref{eq:cq_nshot_POVM_mutual_new}).

Next we look at the second term as follows. Note that 
\[
T_{\hat{\theta}}(u) =\{u': [u'] = [u], \overline{u'} \neq \overline{u} \}.
\]
For any $u' \in T_{\hat{\theta}}(u)$, we have 
\begin{align*}
    \EE_{C} 
\trace &\sparenv{ \parenv{
     (\Pi')^{B'}_{([x], [\Bell])( [u]), (\overline{x},\bar{\Bell}) (u')} }({\rho'})^{B'}_{([x], [\Bell])( [u]), (\overline{x},\bar{\Bell}) (u), \delta }
 } \\
 =& \frac{1}{|\mathcal{L}|^{\zeta(G)+\zeta(H)}} 
\sum_{[x],[\Bell],\overline{x},\bar{\Bell}}
\sum_{\overline{x}',\bar{\Bell}'}
     P([x])
     P(\overline{x})P(\overline{x'}) 
 \trace[ (\Pi')^{B'}_{[x], [\Bell], \overline{x'},\bar{\Bell}'} ({\rho'})^{B'}_{[x], [\Bell], \overline{x},\bar{\Bell}, \delta }] \\
 =& \frac{1}{|\mathcal{L}|^{\zeta(G)}} 
\sum_{[x],[\Bell],\overline{x}',\bar{\Bell'}}
     P([x]) 
     P(\overline{x'}) 
\trace[  (\Pi')^{B'}_{[x], [\Bell], \overline{x'},\bar{\Bell'}} ({\rho'})^{B'}_{[x],[\Bell]}] \\
 \stackrel{(a)}{\leq} &
\sum_{[x],[\Bell],\overline{x}',\bar{\Bell'}}
     \frac{P([x]) 
     P(\overline{x}')}{|\mathcal{L}|^{\zeta(G)}}   \trace[  (\Pi')^{B'}_{[x], [\Bell], \overline{x'},\bar{\Bell'}} 
\mathcal{T}_{[X];[\Bell]} ({\rho}^{B}_{[x]} \otimes 
 |0 \ket \bra 0 |)] + \frac{4 \zeta(G) \delta |B|}{\sqrt{|\mathcal{L}|}} \\
 \stackrel{(b)}{\leq} &  
\sum_{[x],[\Bell],\overline{x}',\bar{\Bell'}}
     \frac{P([x]) 
     P(\overline{x}')}{|\mathcal{L}|^{\zeta(G)}}  
 \trace[( I- (\Pi')^{B'}_{W'_{[x], [\Bell], \overline{x'},\bar{\Bell'}}})  \mathcal{T}_{[X];[\Bell]}({\rho}^{B}_{[x]} \otimes 
 |0 \ket \bra 0 |)] 
  + \frac{4 \zeta(G) \delta |B|}{\sqrt{|\mathcal{L}|}} \\
 \stackrel{(c)}{\leq} & 
\sum_{[x],[\Bell],\overline{x}',\bar{\Bell'}}
     \frac{P([x]) 
     P(\overline{x}')}{ |\mathcal{L}|^{\zeta(G)}} 
\trace[( I- (\Pi')^{B'}_{W'_{[X];[x], [\Bell], \overline{x'},\bar{\mathbf{l'}}}})  \mathcal{T}_{[X];[\Bell]}({\rho}^{B}_{[x]} \otimes 
 |0 \ket \bra 0 |)] + \frac{4 \zeta(G) \delta |B|}{\sqrt{|\mathcal{L}|}} \\
  \end{align*}
 \begin{align*}    
 \stackrel{(d)}{=} & \frac{1}{|\mathcal{L}|^{\zeta(G)}} 
\sum_{[x],[\Bell],\overline{x}',\bar{\Bell'}}
     P([x]) 
     P(\overline{x}') 
\trace[( I- (\Pi)^{\hat{B}}_{W_{[X];[x], \overline{x'}}}) 
 ({\rho}^{B}_{[x]} \otimes 
 |0 \ket \bra 0 |)] + \frac{4 \zeta(G) \delta |B|}{\sqrt{|\mathcal{L}|}} \\
  = & 
\sum_{[x],[\Bell],\overline{x}',\bar{\Bell'}}
     \frac{P([x]) 
     P(\overline{x}')}{|\mathcal{L}|^{\zeta(G)}} 
\trace[(\Pi'')^{B}_{[X];[x], \overline{x'}} ({\rho}^{B}_{[x]} )] + \frac{4 \zeta(G) \delta |B|}{\sqrt{|\mathcal{L}|}} \\
\stackrel{(e)}{\leq} & 2  \left( 2^{-I_{\textnormal{H}}^{\epsilon_{\hat{\theta}}}(\overline{X};[X],B)} \right),
\end{align*}
for large enough $\mathcal{L}$, 
where  (a) follows from 
 (\ref{eq:l_infinity_bound_new}) and
(\ref{eq:l_1_bound_new}),  (b) from  using (\ref{eq:final_POVM_new}),
(c) from using  (\ref{eq:sum_projection_new}), (d) from the fact that $\mathcal{T}_{[X];[\Bell]}$ is an isometry, and hence
\begin{align*}
\trace[ (\Pi')^{B'}_{W'_{[X];[x], [\Bell], \overline{x'},\bar{\Bell'}}} \mathcal{T}_{[X];[\Bell]}({\rho}^{B}_{[x]} \otimes 
 |0 \ket \bra 0 |)] 
=\trace[ (\Pi)^{\hat{B}}_{W_{[X];[x], \overline{x'}}} ({\rho}^{B}_{[x]} \otimes 
 |0 \ket \bra 0 |)],
\end{align*}
and (e) from 
(\ref{eq:cq_nshot_POVM_mutual_new}). 
Combining the three terms we obtain the average probability of error for a fixed $u$ as 
\begin{align*}
   \EE_{C}(P(\mbox{Error}|u)) \leq 
   32\zeta(G)\delta^2 +\frac{8(1+\zeta(G)\delta^2) |\Theta| \epsilon}{\delta^2} 
   + 8 \sum_{\hat{\theta} \in \Theta} |T_{\hat{\theta}}| 2^{-I_{\textnormal{H}}^{\epsilon_{\hat{\theta}}}(\overline{X},[X];B)},
   \end{align*}
for large enough $|\mathcal{L}|$.

\subsection{Proof of Converse for CQ Channel}
\label{app:CQ_Error_C}
We do the preparation for the converse as done in the classical case.
Based on Lemma~\ref{lem:homo_JtoG}, for each group code $\mathds{C}\leq G$, there exists a group $J$ and a homomorphism such that $\mathds{C}$ is the image of the homomorphism. 
Assume now that a group transmission system with parameters $(1, \abs{J}, \epsilon)$ exists over a classical-quantum channel $\cqN = \mathset{\rho_x^B}_{{x\in \cX}}$, where $\mathcal{X}=G$ is an Abelian group, and  $J=\bigoplus_{p\in\mathcal{P}(G)} \bigoplus_{s=1}^{r_p} \mathds{Z}_{p^s}^{k_{p,s}}$. 
{Assume that the homomorphism $\phi$ for the group code $\mathds{C}$ is a one-to-one mapping. }

Let $\hat{\theta}$ be a vector indexed by $(p,s)\in \mathcal{S}(G)$ with $0\leq \hat{\theta}_{p,s}\leq s$.
For a message $u\in J$, construct a one-to-one correspondence between $u_{p,s}\in \mathds{Z}_{p^s}^{k_{p,s}}$ and the tuple $(\tilde{u}_{p,s}, \hat{u}_{p,s})$ where $\tilde{u}_{p,s}\in p^{\hat{\theta}_{p,s}} \mathds{Z}_{p^s}^{k_{p,s}}$ and $ \hat{u}_{p,s} \in \mathds{Z}_{p^{\hat{\theta}_{p,s}}}^{k_{p,s}}$.
Consider the subcode 
$\mathds{C}_1(\hat{\theta}, \hat{u})$
 of $\mathds{C}$ as defined in the classical channel case.
Let $x = \phi(u) +V$ be the channel input and $H_{\hat{\theta}}$ be given as in \eqref{eqn:H_theta}. Then $\mathds{C}_1(\hat{\theta}, \hat{u}) = [x]_{\hat{\theta}} = x+H_{\hat{\theta}}$. That is, there is an one-to-one correspondence between $\hat{U}$  and $[X]_{\hat{\theta}}$. 
Also, $\abs{T_{\hat{\theta}}(u)} = \abs{\mathds{C}_1(\hat{\theta}, \hat{u}) } = \abs{H_{\hat{\theta}}}$ for all $u\in J$.

Define a one-to-one correspondence between $x$ and the tuple $(\tilde{x}_{\hat{\theta}}, [x]_{\hat{\theta}})$ where $\tilde{x}_{\hat{\theta}} = \phi(\tilde{u})$.
Consider a genie-aided receiver which gets access to  $\hat{U}$ and denote it by $\sD^{ga}$. 
Equivalently, this receiver has access to the coset information $[X]_{\hat{\theta}}$ of $X$ and can be realized by a family of POVMs $\mathset{E^{[x]}_x}$. 
Clearly, the average probability of error for this decoder must be not greater than $\epsilon$. Let $X' \in \cX$ be the output of $\sD^{ga}$. 
For every $\hat{\theta} \neq \mathbf{s}$, the average probability of error for this decoder is 
\[
\sum_{\hat{u}} \sum_{x,x'} \Pr(\hat{u}) P_{XX'\mid \hat{U}}(x, x'\mid \hat{u})\one_{ \mathset{x' \neq x}} 
% = \sum_{x,x'} P_{XX'}(x,x')\one_{ \mathset{x' \neq x}}  
\leq \epsilon,
\] where $P_{XX'\mid \hat{U}}(x, x'\vert \hat{u})\triangleq P_{X\mid [X_r]}(x\vert [x]) \trace\sparenv{E^{[x]}_{x'} \rho_x}\,.$ 
% and $[x] = [x]_{\hat{\theta}} = x + \mathds{C}_1(\hat{\theta}, \hat{u}).$

Note that the decoding POVM can be viewed as a CPTP map. This CPTP   maps $\rho^{AB}$ to the (classical) state $P_{XX'}$ denoting the joint distribution of the transmitted codeword $X$ and the decoder's guess $X'$. 
Similarly, it maps $ \rho^{\bar{A}} \rho^{[A]B}$ to $P_{\hat{U}} (P_{X\mid \hat{U}} \otimes P_{X' \mid \hat{U}})$. Hence, it follows from the data processing inequality for $\DHy^\epsilon(\rho\|\sigma)$ that
\begin{align*}
\DHy^\epsilon (P_{XX'}\| P_{\hat{U}} (P_{X\mid \hat{U}} \otimes P_{X' \mid \hat{U}}) ) \le \DHy^\epsilon  (\rho^{AB}\| \rho^{\bar{A}} \rho^{[A]B} ) \ .
\end{align*}

Consider the strategy to distinguish  $P_{XX'}$ and $P_{\hat{U}} (P_{X\mid \hat{U}} \otimes P_{X' \mid \hat{U}})$ as given in the classical channel case, under which the type-I error probability is not larger than $\epsilon$, and the probability of type-II error (misdetection) is $\frac{1}{\abs{H}}$.
That is,
\begin{equation}
\label{eq:subgroup_relativeEntropy_CQ1}
\begin{split}  
\DHy^\epsilon  (\rho^{AB}\| \rho^{\bar{A}} \rho^{[A]B} ) 
&\geq \DHy^\epsilon (P_{XX'}\| P_{\hat{U}} (P_{X\mid \hat{U}} \otimes P_{X' \mid \hat{U}}) )  \\
&= \log_2 \abs{H} = \log_2 \abs{T_{\hat{\theta}}(u)}\, .
\end{split}   
\end{equation}
Using \eqref{eq:T_theta_size}, we may rewrite \eqref{eq:subgroup_relativeEntropy_CQ1} compactly as $    I_{\textnormal{H}}^{\epsilon}(\overline{X}; [X]Y)
\geq (1- \omega_{\hat{\theta}} ) R$. This completes the proof.

\subsection{Proof  of \Tref{thm:CC_nshot_cap}}\label{app:forCCnshot}
\textit{Achievability--}
We use $n$ independent copies of the channel, and make the observation that 
$\mathcal{S}(G)=\mathcal{S}(G^n)$ for all $n \geq 1$. 
Fix an input group $J$ characterized by the corresponding vector $w_{q,s}$, $(q,s) \in \mcs(J)$. 
First we consider the special case where $\alpha$ is the distribution such that 
$\eta = \eta^*$ and $b= b^*$ with probability one, 
% puts all the probability on $\eta^*$ and $b^*$
where $\eta^*$ and $b^*$  are all-zero vectors in their respective spaces
(see the definitions that precede the theorem). 

Using this and Theorem \ref{thm:CC1_achieve}, we see that 
there exists a $(n, \abs{J}, \epsilon')$-code such that 
\[
\epsilon' \leq  \epsilon + \sum_{{\hat{\theta}} \neq \bs }   \exp_2 \bracenv{ 
(1- \omega_{\hat{\theta}} )nR
 -I_{\textnormal{H}}^{\epsilon_{\hat{\theta}}}(\overline{X}^n_{\hat{\theta}}; [X]^n_{\hat{\theta}} Y^n) },\]
where the rate $R$ is given in Equation~\eqref{eqn:rate}, 
and the joint distribution of the input and the output of the channel is given by 
\begin{align*}
P(X^{n}=x^n,Y^n=y^n) &=\prod_{i=1}^{n} P_{X}(x_i) W_{Y|X}(y_i|x_i),
\end{align*}
where $P_X(x)=\frac{1}{|G|}$. 
Then the random vectors will have the following distributions.
For  $x_r^n$ $= (x_{r,1}$, $x_{r,2}, \ldots, x_{r,n})\in G^n$, $[x_r^n]$ denotes the coset  representative of $x_r^n+H^n$ in $G^n$, and the product conditional distribution $P^n_{Y\mid [X]}$ is defined as 
\begin{align*}   
    P^n_{Y\mid [X]}(y^n\,&\vert  \sparenv{x_r^n}) 
    \triangleq \prod_{i=1}^n P_{Y\mid [X]}(y_i\,\vert  \sparenv{x_{r,i}})  
     = \sum_{x^n\in [x_r^n]+H^n} P^n_{X\mid [X]}(x^n| [x_r^n]) W^n_{Y|X}(y^n\vert x^n),
 \end{align*}
 where $P_{[X]}$  and 
 $P_{X\mid [X]}$ are given in Definition~\ref{def:Htheta_cond_CC} and  used in the one-shot case. 
\[
P_{[X]}^n= \frac{|H|^n}{|G|^n}\, , \, P^n_{X\mid [X]}(x^n \,\vert \sparenv{x_r^n}) =
\begin{cases}
    \frac{1}{\abs{H}^n} &\mbox{ if } x^n \in [x_r]^n, \\
    0 &\mbox{ otherwise.} 
\end{cases}
\]
Now since all distributions are in a product form, we can use AEP \cite{wang2012one}
for hypothesis testing relative entropy as: for all $\epsilon>0$, and all $\hat{\theta}$, 
\[
\lim_{n \rightarrow \infty} \frac{1}{n} I_{\textnormal{H}}^{\epsilon, \hat{\theta}}(\overline{X}^n_{\hat{\theta}}; [X]^n_{\hat{\theta}} Y^n) = 
I(\overline{X}_{\hat{\theta}};[X]_{\hat{\theta}},Y)=
I(X;Y|[X]_{\hat{\theta}}).
\]
This gives the desired achievability result in the special case of $\eta^*$ an $b^*$. 

This can be easily extended to the case when $\eta^*$  and $b^*$ are arbitrary vectors. Now the channel input is just restricted to 
$H_{\eta}+b$ instead of the entirety of $G$.
We can apply the above result to this case by defining a new channel that takes inputs in $H_{\eta}$ and then adds $b$ to the input and feeds it to the given channel.  Using the invariance of mutual information under one-to-one mapping of the random variables involved, and the above result, we see that 
the following  rate is achievable
\begin{align*}
R=\min_{\hat{\theta} \neq \bs} \frac{1}{1-\omega_{\hat{\theta}}} I(X_{\eta^*,b^*};Y|[X_{\eta^*,b^*}]_{\hat{\theta}}),
\end{align*}
where $[X_{\eta^*,b^*}]_{\hat{\theta}}=X_{\eta^*,b^*}+
H_{\eta^*+\hat{\theta}}$.

Toward extending these arguments to the general $\alpha$, we note that the random homomorphism used by the encoder has the structure that for a given message, the outputs corresponding to different $\mathbb{Z}_{p^r}$ components of $G$ are mutually independent. Consider a set of positive integers
$n_{\eta,b}$ indexed by $\eta,b$, such that 
$\alpha_{\eta,b}=n_{\eta,b}/n$, for large $n$, where  $n=\sum_{\eta,b} n_{\eta,b}$.  We construct random encoder such that its output takes values in 
$H_{\eta}+b$ for $n_{\eta,b}$ samples for all $\eta,b$.  

Next we note the following superadditivity of smoothed Hypothesis testing relative entropy:
\begin{align}
D_{\textnormal{H}}^{\epsilon}(P_1 \times P_2 \| Q_1 \times Q_2) &\stackrel{(a)}{\geq} -\log_2 [Q_1(A_1) \times Q_2(A_2)] \\ &= D_{\textnormal{H}}^{\epsilon/2}(P_1 \|Q_1)+D_{\textnormal{H}}^{\epsilon/2}(P_2 \|Q_2),
\label{eq:superadditive}
\end{align}
where (a) follows by defining $A_i$ to be the decision region that achieves the optimality in $D_{\textnormal{H}}^{\epsilon/2}(P_i \| Q_i)$, and noting that 
$P_1(A_1)P_2(A_2)=1-\epsilon+\epsilon^2/4 \geq 1-\epsilon$. 
Using such a suboptimal decision region that has a product structure across different values of $\eta$ and $b$, we see that 
there exists a $(n, \abs{J}, \epsilon')$-code such that 
\[
\epsilon' \leq  \epsilon + \sum_{{\hat{\theta}} \neq \bs }   \exp_2 \bracenv{ 
(1- \omega_{\hat{\theta}} )nR
 -\sum_{\eta,b} \alpha_{\eta,b} I_{\textnormal{H}}^{\epsilon_{\hat{\theta}}}(\overline{X}^{n_{\eta,b}}_{\hat{\theta}}; [X]^{n_{\eta,b}}_{\hat{\theta}} Y^{n_{\eta,b}}) }.\]
Using the AEP, we see that the following rate is achievable:
\begin{align*}
\min_{\hat{\theta} \neq \bs}
\frac{1}{1-\omega_{\hat{\theta}}} \sum_{\eta,b} \alpha_{\eta,b} I(X_{\eta,b};Y|[X_{\eta,b}]_{\hat{\theta}}),
\end{align*}
for rational $\alpha_{\eta,b}$. 
By approximating  and optimizing over $\alpha$ and the input group, we get the desired result.

\textit{Converse--}
 The converse follows from \cite{sahebi2015abelian}.

\subsection{Proof of \Tref{thm:CQ_nshot_cap}}
\label{app:forCQnshot}

\textit{Achievability}-- The achievability follows by exploiting the product nature of the input distribution, the quantum Stein's lemma \cite[Theorem 2]{ogawa2000strong}, the superadditivity of smoothed hypothesis testing quantum relative entropy, and Theorem \ref{thm:CQ1_achieve_1} as in the classical case, We skip the details for conciseness. 

\textit{Converse}-- Toward a converse, consider the following arguments.  Consider an arbitrary shifted group code $\mathds{C}$ with parameters
$n,k$ and $w$.  We assume that the associated
homomorphism is a one-to-one mapping.    Recall that we can express the code compactly as follows:
\begin{align*}
\mathds{C} &= \Bigg\{ \bigoplus_{i=1}^n \left[ \bigoplus_{(p,r,m)=\mcg(G)} \sum_{s=1}^{r_p} a_{p,s} g^{(i)}_{(p,s) \rightarrow (r,m)} 
+B^{(i)} \right]  
  :  a_{p,s} \in \mathds{Z}_{p^s}^{k w_{p,s}}, \forall
(p,s) \in \mcs(G) \Bigg\}.
\end{align*}

For every pair of vectors $(\eta,b)$ as defined earlier, define 
\begin{align*}
\Gamma_{\eta,b} = \big\{ i\in [1,n]: 
g^{(i)}_{(p,s) \rightarrow
    (r,m)} \in p^{\eta_{p,r,m,s}+|r-s|^+} \mathds{Z}_{p^r}^{kw_{p,s}}, g^{(i)}_{(p,s) \rightarrow
    (r,m)} \not\in &p^{\eta_{p,r,m,s}+1+|r-s|^+}
  \mathds{Z}_{p^r}^{kw_{p,s}}, \\
  &B^{(i)}=b, \forall (p,r,m,s)  \big\}
\end{align*}

Let $\hat{\theta}$ be an arbitrary vector whose components 
$\hat{\theta}_{p,s}$ are indexed by $(p,s) \in \mcs(G)$ and 
satisfy $0 \leq \hat{\theta}_{p,s} \leq s$.  Construct a one-to-one correspondence $a_{p,s} \leftrightarrow 
(\tilde{a}_{p,s},\hat{a}_{p,s})$ where $\tilde{a}_{p,s} \in p^{\hat{\theta}_{p,s}} \mathds{Z}_{p^s}^{k w_{p,s}}$ and 
 $\hat{a}_{p,s} \in \mathds{Z}_{p^{\hat{\theta}_{p,s}}}^{k w_{p,s}}$.
For an arbitrary $\hat{a}_{p,s} \in \mathds{Z}_{p^{\hat{\theta}_{p,s}}}^{k w_{p,s}}$, 
consider the following subcode of $\mathds{C}$:
\begin{align*}
\mathds{C}_1(\hat{\theta},\hat{a}) \!\! &= \!\! \Bigg\{
\bigoplus_{i=1}^n \! \bigoplus_{(p,r,m)=\mcg(G)} \! 
\sum_{s=1}^{r_p} (\hat{a}_{p,s}+\tilde{a}_{p,s}) g^{(i)}_{(p,s) \rightarrow (r,m)} 
+B  : \tilde{a}_{p,s} \in p^{\hat{\theta}_{p,s}}
\mathds{Z}_{p^s}^{k w_{p,s}}, \forall (p,s) \in \mcs(G) \Bigg\}.
\end{align*}
The rate of the code $\mathds{C}_1(\hat{\theta},\hat{a})$
 is given by 
$(1-\omega_{\hat{\theta}})\frac{k}{n} \sum_{(p,s) \in \mcs(G)} s w_{p,s} \log p$.

For a given CQ channel $\mathcal{N}$, suppose that rate $R$ is achievable using group codes. Consider an
arbitrary $\epsilon>0$. This implies that 
there exists a shifted group code 
$\mathds{C}$  with parameters $n,k$ and $w$ that  yields  an average error probability $\tau$ such that $\tau \leq  \epsilon$ and $\frac{k}{n}\sum_{(p,s) \in \mcs(G)} s w_{p,s}
\log p\geq R- \epsilon$.    
Using a uniform
distribution on $a$,  we let  $X_i$ denote the random channel input at the $i$th channel use induced by this code. 

\begin{figure*}[!t]
%\vspace*{4pt}
%\setcounter{mytempeqncnt}{\value{equation}}
%\setcounter{equation}{141}
\begin{align}
\mathds{C}_2(\hat{\theta}) &=\left\{ \bigoplus_{\eta,b} \bigoplus_{i \in \Gamma(\eta,b)} 
\left[ \bigoplus_{(p,r,m)=\mcg(G)} 
\sum_{s=1}^{r_p} a_{p,s} g^{(i)}_{(p,s) \rightarrow (r,m)} 
+b +H_{\eta+\hat{\theta}} \right] : a_{p,s} 
\in  \mathds{Z}_{p^s}^{k w_{p,s}}, \forall (p,s) \in \mcs(G) \right\}  \nonumber \\
&\stackrel{(a)}{=}
\left\{ \bigoplus_{\eta,b} \bigoplus_{i \in \Gamma(\eta,b)} 
\left[ \bigoplus_{(p,r,m)=\mcg(G)} 
\sum_{s=1}^{r_p} \hat{a}_{p,s} g^{(i)}_{(p,s) \rightarrow (r,m)} 
+b +H_{\eta+\hat{\theta}} \right] : \hat{a}_{p,s} 
\in  \mathds{Z}_{p^{\hat{\theta}_{p,s}}}^{k w_{p,s}}, \forall (p,s) \in \mcs(G) \right\} \nonumber  
\end{align}
\hrulefill
\end{figure*}

Using the fact that  $\tilde{a}$ is uniformly distributed over its range, for $i \in \Gamma_{\eta,b}$, in the code 
$\mathds{C}_1(\hat{\theta},\hat{a})$, the channel input $X_i(\hat{\theta},\hat{a})$ 
at the $i$th channel use has the following distribution
\[
P(X_i(\hat{\theta},\hat{a})=\beta)=
\prod_{(p,r,m) \in \mcg(G)}  \!\!\!
p^{-|r-\pmb{\theta}(\eta+\hat{\theta})_{(p,r,m)}|^+} = \frac{1}{|H_{\eta+\hat{\theta}}|},
\]
if 
\[
\beta_{(p,r,m)} \in  \sum_{s=1}^{r_p} \hat{a}_{p,s} g^{(i)}_{(p,s) \rightarrow (r,m)} +b_{p,r,m}+ p^{\pmb{\theta}(\eta+\hat{\theta})_{(p,r,m)}} 
\mathds{Z}_{p^r}
\]
 for all $(p,r,m) \in \mcg(G)$, and $P(X_i(\hat{\theta},\hat{a})=\beta)=0$  otherwise. 

Let $A$ denote the random message of the group
transmission system of the code $\mathbb{C}$.  
Let $\hat{A}$ denote the part of the random message such that  
 $\hat{A}_{p,s} \in \mathds{Z}_{p^{\hat{\theta}_{p,s}}}^{k w_{p,s}}$,
 for all $(p,s) \in \mathcal{S}(G)$.   Now consider a genie-aided
 receiver which gets access to $\hat{A}$. This receiver implements the POVM as in the original case, but substitutes its estimate of $\hat{A}$  with the actual realization given by the genie. The probability of error of this receiver is clearly no greater than $\tau$. 
 Using the definition of quantum mutual information
$I(X;Y | [X]_{\hat{\theta}})
 = D(\rho^{A B} || \rho^{\bar{A}_{\hat{\theta}} } \rho^{[A]_{\hat{\theta}} B} )$,
 classical Fano's inequality on the outcome of the POVM (measurement) \cite[Chapter 5]{holevo2019quantum}, quantum data processing inequality by viewing POVM as a CPTP map, we have for every $\hat{\theta}$ with $0 \leq \hat{\theta}_{p,s} \leq s$, we have the following argument with regard to quantum mutual information:
\begin{align}
(1-\omega_{\hat{\theta}})(R-\epsilon)(1-\tau) -\frac{1}{n}  &\leq
\frac{1}{n} \sum_{\hat{a}} P(\hat{a})
I(X^n(\hat{\theta},\hat{a});Y^n|\hat{A}=\hat{a}) \nonumber \\
&\leq \sum_{\hat{a}} P(\hat{a}) \frac{1}{n} \sum_{i=1}^n
I(X_i(\hat{\theta},\hat{a});Y_i)  \nonumber \\
&\stackrel{(a)}{=}  \sum_{\eta,b} \sum_{i \in \Gamma(\eta,b)}
\!\! \sum_{\hat{a}} \! \frac{P(\hat{a}) I(X_i;Y_i|X_i \in
  \hat{a}\pmb{g}^{(i)}+b+H_{\eta+\hat{\theta}} )}{n} \nonumber \\ 
&\stackrel{(b)}{=} \sum_{\eta,b} \sum_{i \in \Gamma(\eta,b)} \frac{1}{n}  I(X_i;Y_i|X_i \in \hat{A}\pmb{g}^{(i)}+b+H_{\eta+\hat{\theta}} ) \\
&\stackrel{(c)}{=} \sum_{\eta,b} \frac{|\Gamma(\eta,b)|}{n}
  I(X_{\eta,b};Y|[X_{\eta,b}]_{\hat{\theta}}),
\end{align}
where in (a) we have expressed 
$\bigoplus_{(p,r,m) \in \mcg(G)} \sum_{s=1}^{r_p} \hat{a}_{p,s} g^{(i)}_{(p,s) \rightarrow (r,m)} +b_{p,r,m}$ as
$\hat{a} \pmb{g}^{(i)}+b$, in (b) $\hat{A}$ denotes the random variable corresponding to $\hat{a}$, 
in (c) we have used the fact that $\hat{A} \pmb{g}^{(i)}+b+H_{\eta+\hat{\theta}}$ is uniform over the set
 of cosets of $H_{\eta+\hat{\theta}}+b$ in $H_{\eta}+b$. Hence the converse follows. 

\endproof

%% Or you use manual references (pay attention to consistency and the
%% formatting style!):
\end{document}